\documentclass[aps,prd,reprint,twocolumn,superscriptaddress,nofootinbib,longbibliography]{revtex4-2}

 \usepackage{placeins}
 \usepackage{adjustbox}
 \usepackage{hyperref}

\usepackage{amsmath,amssymb,bm,booktabs,graphicx}
\usepackage{float}
\hypersetup{colorlinks=true,citecolor=blue,linkcolor=blue,urlcolor=blue}
\graphicspath{{figures/}}

\usepackage{xcolor}

\begin{document}

\title{Constrained field-space estimators for background-field lattice QCD observables}

\author{Scott Moerschbacher}
\email{scott@openinquiryresearch.org}
\affiliation{Open Inquiry Research Group, Monrovia, MD 21770, USA}

\author{Frank X. Lee}
\email{fxlee@gwu.edu}
\affiliation{Physics Department, The George Washington University, Washington, DC 20052, USA}

\date{\today}

\begin{abstract}
Background-field lattice calculations extract weak-field response coefficients from correlators computed at several field strengths. We treat this extraction as linear estimation in field space and compare fixed cancellation stencils with covariance-optimized, generalized-least-squares estimators of the same coefficient, using archival nucleon double ratios on about 1000 quenched configurations at six quark masses and decuplet correlators for which exact answers follow from the lattice contractions. The fixed $O(B^4)$--$O(B^8)$ stencils are nearly coincident and cost almost nothing in precision. Weight-free null-space tests reject the two-term response model in most nucleon ensembles; whether they reject the three-term model depends on the temporal fits and on the noise of the estimated temporal covariance, which raises the null-contrast error by a factor of 1.7--2.4. With that noise included, covariance optimization gains 1--2\% in precision at the five lighter masses (3--30\% with two-state time fits, which also lower moments by 3--10\%) and nothing on independent data halves, while its weights carry three to four orders of magnitude more leverage on omitted response modes than the fixed stencils and move the moments by up to 1.8 marginal standard errors. The decuplet controls show what such leverage does under misspecification: the first-order responses obey $\mu_{\Delta^+}/\mu_{\Delta^{++}}=\tfrac12$ and $\mu_{\Delta^0}=0$ exactly. The lowest-order estimate misses the first by 10--25 standard errors. The fixed $O(B^4)$--$O(B^8)$ stencils satisfy it within errors in the production windows, whereas the correlated linear fit in $B^2$, with the same constraints as the $O(B^4)$ stencil, misses it by 3--17\%; the null-space test rejects that fit's model. The results distinguish statistical precision from robustness to omitted response modes; they are not physical-point magnetic moments.
\end{abstract}

\maketitle

\section{Introduction}

The background-field method determines electromagnetic properties of hadrons from the response of lattice correlation functions, and hence extracted energies, to applied external electromagnetic fields \cite{Martinelli1982,Bernard1982}. Early calculations used progressively weak fields to isolate magnetic moments and polarizabilities, with opposite field orientations employed to separate odd and even response components \cite{Lee2005MagPol,Lee2005BaryonMom,Lee2008MagMom,Alexandru2009BFM}. Modern implementations retain the same basic response logic while improving field implementation, state isolation, and correlated fitting \cite{Detmold2010Electric,Detmold2011Review,Davoudi2015Background,Bignell2020Pion,Kabelitz2024Octet}.

At finite field strength the desired weak-field coefficient is accompanied by higher-order response. For an odd magnetic response one may write
\begin{equation}
\frac{E(B)-E(-B)}{2B}=c_1+c_3B^2+c_5B^4+\cdots,
\label{eq:oddreduced}
\end{equation}
while an analytic even response gives
\begin{equation}
\frac{E(B)+E(-B)-2E(0)}{2B^2}=c_2+c_4B^2+c_6B^4+\cdots.
\label{eq:evenreduced}
\end{equation}
Higher-order terms are a familiar finite-field systematic. Previous work has addressed them by using sufficiently weak fields, comparing subsets of field strengths, or including higher powers in response fits \cite{Lee2005MagPol}. Several modern calculations likewise employ multiple matched fields and exploit same-configuration correlations in ratios and correlated fits \cite{Primer2014,Bignell2017Nucleon,Bignell2020Pion,Kabelitz2024Octet}.

Here we reorganize this established extraction problem from a numerical-analysis perspective. The desired response coefficient is a derivative, or equivalently a coefficient of a specified weak-field basis, evaluated at zero field. Measurements at several field strengths may therefore be regarded as samples in \emph{field space}. Linear combinations of those samples can be chosen to preserve the desired coefficient while annihilating specified nuisance modes.

This viewpoint makes field redundancy explicit. Additional field values can be used to impose further response-mode constraints. If more field values are available than required by those constraints, the remaining estimator degrees of freedom can instead be used for statistical optimization. The balance between these uses is not fixed a priori and may depend on ensemble statistics, cross-field covariance, and the size of unresolved response modes.

The mathematics is standard. Fixed cancellation stencils are polynomial interpolation, the optimized estimators are constrained generalized least squares (with four fields, the O4-constrained one is the correlated linear fit in $B^2$ to all fields), and the null-space tests are residual tests of a linear model. We do not claim a new estimator. Our contribution is empirical and diagnostic: (i) a systematic comparison, on real background-field correlators, of fixed and covariance-optimized estimators of the same weak-field coefficient, including their behavior on independent data and under the noise of the estimated covariances; (ii) controls derived from the lattice contractions that supply known targets against which estimator displacement can be measured; and (iii) a synthetic study with known covariance and response coefficients that separates estimator variance, covariance-estimation noise and omitted-mode displacement. Writing the extraction in field space makes these comparisons uniform: the cancellation constraints, the remaining statistical directions and their sensitivity to omitted response modes appear explicitly. Throughout, O2, O4, O6 and O8 denote stencils whose leading uncancelled analytic contamination is $O(B^2)$, $O(B^4)$, $O(B^6)$ and $O(B^8)$ in the reduced response, not the degree of a fitted polynomial, and we distinguish three properties of an estimator: minimum variance, statistical stability under resampling, and systematic stability against unconstrained response modes. Optimizing one does not in general optimize the others.

Section~\ref{sec:constrained} develops the constrained formulation and Sec.~\ref{sec:leverage} the role of covariance optimization and omitted-mode leverage. Section~\ref{sec:methods} describes the lattice data and Sec.~\ref{sec:fixedresults} the fixed estimator hierarchy. Section~\ref{sec:blue_results} presents the null-space specification tests and their calibration, Sec.~\ref{sec:blue} the covariance-optimized estimators, including a synthetic study, and Sec.~\ref{sec:decuplet} the known-answer decuplet controls. The appendices document the archival reconstruction, the decuplet identities at the level of the correlators, and a charged-state example.

\section{Constrained estimators in field space}
\label{sec:constrained}

\subsection{General formulation}

Let a reduced field response be written as
\begin{equation}
y(B)=\theta+\sum_{k=1}^{K-1}a_k\phi_k(B)+r(B),
\label{eq:generalresponse}
\end{equation}
where $\theta$ is the desired coefficient, $\phi_k(B)$ are response
modes to be removed, and $r(B)$ contains modes not explicitly constrained.
Given measurements $y_i=y(B_i)$, consider the linear estimator
\begin{equation}
\widehat{\theta}
=
\sum_i w_i y_i
=
\bm w^T\bm y.
\label{eq:linear_estimator}
\end{equation}
Substituting Eq.~(\ref{eq:generalresponse}) gives
\begin{equation}
\widehat{\theta}
=
\theta\sum_i w_i
+
\sum_{k=1}^{K-1}a_k\sum_i w_i\phi_k(B_i)
+
\sum_i w_i r(B_i).
\label{eq:constraint_expansion}
\end{equation}
The cancellation conditions are therefore explicit. Preserving the
target coefficient requires
\begin{equation}
\sum_i w_i=1,
\label{eq:preserve_target}
\end{equation}
while annihilating each selected nuisance mode requires
\begin{equation}
\sum_i w_i\phi_k(B_i)=0,
\qquad k=1,\ldots,K-1.
\label{eq:annihilate_modes}
\end{equation}

These conditions may be collected in matrix form as
\begin{equation}
A^T\bm w=\bm b,
\qquad
\bm b=(1,0,\ldots,0)^T,
\label{eq:constraints}
\end{equation}
where the columns of $A$ are the sampled constant mode and nuisance
modes. For example, preserving the constant term while cancelling
$B^2$ and $B^4$ response modes corresponds to
\begin{equation}
A=
\begin{pmatrix}
1 & B_1^2 & B_1^4\\
1 & B_2^2 & B_2^4\\
\vdots & \vdots & \vdots\\
1 & B_N^2 & B_N^4
\end{pmatrix},
\qquad
\bm b=
\begin{pmatrix}
1\\
0\\
0
\end{pmatrix}.
\label{eq:example_design}
\end{equation}
Thus $A^T\bm w=\bm b$ is precisely the compact form of the
normalization and cancellation conditions above.

This formulation does not require a monomial basis. The response
functions included in $A$ should reflect the physics of the observable;
polynomial stencils are one important special case.

If $\Sigma$ denotes the covariance matrix of the field responses, the
minimum-variance estimator within a given constraint family is obtained
from
\begin{equation}
\min_{\bm w}\ \bm w^T\Sigma\bm w
\qquad\text{subject to}\qquad
A^T\bm w=\bm b.
\label{eq:constrained_minimization}
\end{equation}
For nonsingular $\Sigma$ and full-rank constraints, the resulting best
linear unbiased estimator (BLUE) has weights
\begin{equation}
\bm w_{\rm BLUE}
=
\Sigma^{-1}A
\left(A^T\Sigma^{-1}A\right)^{-1}\bm b.
\label{eq:blueweights}
\end{equation}
The matrix $A$ therefore specifies which response directions are
preserved or annihilated, while $\Sigma$ determines the
minimum-variance choice among all weights satisfying those constraints.

\subsection{Odd and even analytic responses}

For the odd response, Eq.~(\ref{eq:oddreduced}), the reduced response is $y(B)=[E(B)-E(-B)]/(2B)=c_1+c_3B^2+c_5B^4+\cdots$; for the analytic even response, Eq.~(\ref{eq:evenreduced}), it is $z(B)=[E(B)+E(-B)-2E(0)]/(2B^2)=c_2+c_4B^2+c_6B^4+\cdots$. The field-space algebra is identical in the two cases: the target is the constant term of a reduced response whose analytic contamination is organized in powers of $B^2$. Writing $B_i=m_iB$ and $x_i\equiv m_i^2$, preserving the target and removing successive powers of $B^2$ requires
\begin{equation}
\begin{gathered}
\sum_i w_i = 1,\qquad
\sum_i w_i x_i = 0,\\
\sum_i w_i x_i^2 = 0,\qquad
\sum_i w_i x_i^3 = 0.
\end{gathered}
\label{eq:sum_constraint}
\end{equation}
The O2 estimator imposes only the first condition, O4 the first two, O6 the first three and O8 all four, so that the leading residual analytic contamination is $O(B^2)$ for O2, $O(B^4)$ for O4, $O(B^6)$ for O6 and $O(B^8)$ for O8. For the field magnitudes used here, $m_i=(1,2,4,8)$ and $x_i=(1,4,16,64)$, $K$ conditions leave $4-K$ weight directions free; the fixed hierarchy is the special choice that satisfies them using only the $K$ weakest fields,
\begin{align}
\widehat\theta^{(2)} &= y(B),\nonumber\\
\widehat\theta^{(4)} &= \tfrac43y(B)-\tfrac13y(2B),\nonumber\\
\widehat\theta^{(6)} &= \tfrac{64}{45}y(B)-\tfrac49y(2B)+\tfrac1{45}y(4B),\nonumber\\
\widehat\theta^{(8)} &= \tfrac{4096}{2835}y(B)-\tfrac{64}{135}y(2B)+\tfrac4{135}y(4B)-\tfrac1{2835}y(8B),
\label{eq:fixedhierarchy}
\end{align}
where $y$ may denote either reduced response. O4 and O6 are one and two Richardson extrapolation steps in $B^2$ \cite{Richardson1927}.

\subsection{Relation to polynomial fits}

The fixed estimators are algebraically related to conventional response fits. Fitting $y(B)$ and $y(2B)$ exactly to $\theta+aB^2$ and evaluating the intercept gives the $O(B^4)$ stencil above; exact three- and four-point interpolation gives the $O(B^6)$ and $O(B^8)$ stencils. Thus, when the number of field values equals the number of retained basis terms, the fixed stencil and an exact polynomial-intercept extraction are algebraically equivalent.

The distinction made here is operational rather than algebraic. The constrained formulation exposes the nuisance-mode annihilation conditions, makes the $n-K$ remaining estimator degrees of freedom for $n$ fields explicit, and permits those directions to be allocated independently to further systematic cancellation or statistical optimization; what it adds is the diagnostics this makes available, not a new representation of the polynomial.

With more fields than retained terms, a correlated (generalized-least-squares) polynomial fit with the same basis is the minimum-variance member of the constrained family, Eq.~(\ref{eq:blueweights}): with four fields, the correlated linear fit in $B^2$ is O4-constrained BLUE and a correlated constant fit is BLUE with normalization only. For a well-resolved response with a justified fit basis, such a fit remains fully appropriate. The field-space hierarchy offers a complementary diagnostic: rather than estimating a nuisance coefficient and deciding whether it is needed, one constructs an estimator whose response to that mode is exactly zero and compares it, with full correlation retained, to estimators with fewer or additional cancellations. Successive differences such as $\widehat\theta^{(4)}-\widehat\theta^{(2)}$ therefore measure the effect of removing a specified response mode directly.

\section{Redundancy, covariance, and response-mode leverage}
\label{sec:leverage}

With $n$ field values subject to $K$ independent constraints, $n-K$ estimator directions remain free; for four fields, O2, O4, O6 and O8 leave three, two, one and none.
If the cross-field covariance $\Sigma$ were known, the constrained minimum-variance weights would be those of Eq.~(\ref{eq:blueweights}), the standard constrained GLS/BLUE solution \cite{ValassiChierici2014}. They minimize $\bm w^T\Sigma\bm w$ subject to the imposed constraints, but they do not control response modes omitted from $A$. If the true reduced response contains an additional mode $a_\star\phi_\star(B)$, its contribution to the estimator is
\begin{equation}
\delta\widehat\theta_\star
=a_\star\bm w^T\bm\phi_\star,
\qquad
(\bm\phi_\star)_i=\phi_\star(B_i).
\label{eq:leverage}
\end{equation}
The scalar $\bm w^T\bm\phi_\star$ therefore measures the estimator's leverage on that omitted response mode. A statistically stable minimum-variance solution can still be systematically fragile if this leverage is large. This is the mechanism behind Peelle's pertinent puzzle \cite{DAgostini1994}: a generalized-least-squares fit to strongly correlated data with an imperfect model can return an estimate far outside the range of the data.

In practice $\Sigma$ is also estimated from a finite ensemble. Thus two distinct questions arise: whether the covariance directions exploited by optimized weights generalize out of sample, and whether the resulting weights remain well conditioned against physically plausible omitted response modes. The archival tests below separate these questions.

For four fields the displacement between any two O4-constrained estimators can be written exactly. Interpolating the measured responses as $y(x)=\alpha+\beta x+\gamma x^2+\delta x^3$, with $x_i=(1,4,16,64)$, and defining $L_4(\bm w)=\bm w^T\bm x^2$ and $L_6(\bm w)=\bm w^T\bm x^3$, any two weights $\bm w,\bm v$ that preserve the constant and annihilate $x$ satisfy
\begin{equation}
\bm w^T\bm y-\bm v^T\bm y=[L_4(\bm w)-L_4(\bm v)]\gamma+[L_6(\bm w)-L_6(\bm v)]\delta .
\label{eq:leverageidentity}
\end{equation}
This is an interpolation identity, not evidence that the physical response truncates at cubic order: it identifies which directions of the measured response produce a displacement, not the bias relative to the unknown zero-field coefficient. The fixed O4 stencil has $L_4=-4$ and $L_6=-20$. The size of $L_4$ and $L_6$ depends on the normalization of $x$, here the square of the weakest field: with $x$ measured in units of the strongest field they would be smaller by $64^2$ and $64^3$, and $\gamma$ and $\delta$ larger by the same factors. Only the products, the displacements, are independent of the normalization. We therefore call a leverage large only relative to the fixed stencil of the same order in the same normalization, and quote its consequences as displacements.

\section{Lattice data and analysis}
\label{sec:methods}
\subsection{Ensemble and field implementation}

The correlators were computed on quenched configurations generated with the Wilson gauge action at $\beta=6.0$ on a $24^3\times48$ lattice, the ensemble of Ref.~\cite{AlexandruLee2010}, with Wilson valence fermions. We use $a^{-1}=2.116$~GeV ($a\simeq0.093$~fm), consistent with the Sommer scale at $\beta=6.0$ \cite{NeccoSommer2002} for $r_0\simeq0.50$~fm. The main sample, $\mathrm{poly11\mbox{-}small}$, contains 992 proton and 999 neutron correlators at $\kappa=(0.1567,\allowbreak 0.1566,\allowbreak 0.1565,\allowbreak 0.1564,\allowbreak 0.1562,\allowbreak 0.1546)$, corresponding to $m_\pi\simeq(338,\allowbreak 362,\allowbreak 384,\allowbreak 405,\allowbreak 444,\allowbreak 693)$~MeV (plateau averages of the weakest-field effective masses give $M_N\simeq1150$--1480~MeV and $M_\Delta\simeq1410$--1650~MeV, an $N$--$\Delta$ splitting of 260~MeV at the lightest and 160~MeV at the heaviest mass), together with decuplet correlators for the $\Delta^{++}$, $\Delta^+$, $\Delta^0$ and $\Delta^-$ (999, 892, 871 and 901 configurations); Sec.~\ref{sec:decuplet} uses all four. Smaller, distinct samples enter only where stated: a 90-configuration proton sample ($\mathrm{poly0}$) at $\kappa=(0.1558,\allowbreak 0.1549,\allowbreak 0.1539,\allowbreak 0.1530,\allowbreak 0.1520,\allowbreak 0.1479)$ and a 100-configuration sample at the same quark masses with the $\mathrm{poly11}$ origin, both used only to validate the reconstruction and the origin dependence (Appendix~\ref{app:archive}). The samples are not pooled. Appendix~\ref{app:archive} describes how the archived correlators and the original analysis chain were reconstructed and validated.

The quark source is at $(11,11,11,0)$. A uniform magnetic field along $z$ acts on the valence quarks only, through the Abelian phase $\exp[i\,q\,\eta\,(x-x_0)]$ on each $y$-directed link, where $q$ is the quark charge in units of $q_d$ ($q=1$ for $d$, $-2$ for $u$) and $x_0=11$, and $|\eta|=\eta_0(1,2,4,8)$ with $\eta_0=0.001364$. With $\eta=q_dBa^2$ and $q_d=-e/3$, the field magnitudes are $|eB|a^2=3|\eta|$. In the valence Dirac operator the links that would cross the $x$, $y$ and $t$ boundaries are set to zero before the phases are applied, while $z$ stays periodic; the gauge configurations themselves are periodic and unchanged. The valence quarks therefore propagate in a $24\times24$ open box in the $x$--$y$ plane, and only forward in time from the source at $t=0$. Because no link closes the $x$ or $y$ direction, the usual flux-quantization condition of a periodic torus does not apply and every plaquette in the $x$--$y$ plane carries the same flux; we describe the boundaries as removed boundary-crossing links rather than assuming textbook Dirichlet conditions. Moving the gauge origin $x_0$ is then an exact symmetry of the quark action: neutral correlators are unchanged, while the zero-momentum projection of a charged correlator acquires a momentum shift (Appendix~\ref{app:archive}). The magnetic moments agree within errors between origins (Appendix~\ref{app:archive}).

The relevant length scales are the magnetic length $\ell_B=(|e|B)^{-1/2}$ and the distance from the source to the removed boundary links, 11--12 sites. For $f=1,2,4,8$, $\ell_B/a=15.6$, 11.1, 7.8 and 5.5 (the quark-level lengths $(|e_q|B)^{-1/2}$ are larger by $\sqrt3$ for $d$ and $\sqrt{3/2}$ for $u$): $\ell_B$ exceeds the source-to-boundary distance only at the weakest field, is comparable to it at $2B$, and falls below it at $4B$ and $8B$, so for a charged state the transverse confinement crosses over from the walls to the field within the range of fields used. Departures from a low-order field expansion that appear mainly through $8B$ may therefore reflect boundary or finite-volume effects as well as higher-order electromagnetic response. The latter is also expected there: in physical units $\ell_B\simeq1.46$, 1.03, 0.73 and 0.52~fm, so at $8B$ the magnetic length is comparable to or smaller than the nucleon itself, and the weak-field expansion, roughly an expansion in $|e|Br^2=(r/\ell_B)^2$ with $r$ the hadron size, is least reliable there. For charged states such as the proton, Landau-level effects add to these \cite{TiburziVayl2013,Bignell2020Nucleon}. We do not attempt an infinite-volume interpretation of the field coefficients; all statements below concern estimators applied to these data.

\subsection{Magnetic double ratios}

For positive-parity nucleon correlators the two real spin components define the matched-field double ratio
\begin{equation}
R(t,B)=\frac{G_{11}(+B)G_{22}(-B)}{G_{11}(-B)G_{22}(+B)}.
\label{eq:doubleratio}
\end{equation}
For each field magnitude, we fit $R(t,B)=A\exp[-m(B)t]$, with $t$ in lattice units and $m(B)$ a dimensionless exponent, using the correlated temporal covariance. With spin-component energies $E_\pm=M\pm\mu B+\cdots$, the spin--field double ratio gives $m(B)=4a\mu B+\cdots$. With $\eta=q_dBa^2$, $q_d=-e/3$, and the sign mapping of the archived field tags (Appendix~\ref{app:archive}), the reduced response is therefore
\begin{equation}
\begin{gathered}
y(B)=\frac{m(B)}{12\eta}=\frac{\mu(B)}{ea},\\
\frac{\mu(B)}{\mu_N}=2M_p a\,y(B)
=\frac{2(0.938\,{\rm GeV})}{2.116\,{\rm GeV}}\,y(B)\\
=0.88658\,y(B).
\end{gathered}
\label{eq:magresponse}
\end{equation}
The simulated nucleon mass cancels from this conversion. Equation~(\ref{eq:magresponse}) differs by an overall factor from the normalization used in the original analysis notebooks (Appendix~\ref{app:archive}); the factor leaves stencil weights, standard-error ratios, specification tests and leverage diagnostics unchanged. The same aligned configuration deletion is used at every field. The production fit windows are $t=12$--20 for the proton and $t=17$--26 for the neutron (Sec.~\ref{sec:windows}).

The double ratio keeps the two spin components separate, and this matters for any field-even nucleon analysis built from the same correlators. At fixed field the spin-averaged correlator $[G_{11}(B)+G_{22}(B)]/2$ is a sum of two exponentials with energies $E\pm\mu B$; for equal overlaps its effective energy is
\begin{equation}
E_{\rm eff}(t)=E-\mu B\tanh(\mu B t)+\cdots\simeq E-(\mu B)^2 t,
\label{eq:spinavg}
\end{equation}
a term that is even in $B$, proportional to $B^2$ at weak field, and grows linearly with Euclidean time. Averaging over $\pm B$ does not remove it, so a single-exponential fit to the spin-averaged correlator mimics a field-even response. Field-even nucleon responses must therefore be extracted from spin-resolved energies. The magnetic-moment analysis below is unaffected, because the double ratio, Eq.~(\ref{eq:doubleratio}), cancels field-even energy shifts identically.

\subsection{Jackknife errors}

For aligned delete-one estimates $\theta_i$ from $N$ configurations, we use
\begin{equation}
\mathrm{Cov}_{\rm JK}(\theta,\varphi)=\frac{N-1}{N}
\sum_{i=1}^N(\theta_i-\bar\theta)(\varphi_i-\bar\varphi).
\label{eq:jkcov}
\end{equation}
The full high-statistics analysis uses nonlinear correlated exponential fits in the delete-one replicas. For each field, the temporal covariance of the double ratio is estimated once from the delete-one samples and held fixed in every replica fit; it is not re-estimated inside each replica. Holding the temporal covariance fixed omits the noise of its estimate, which also affects the $\chi^2$ statistics of correlated fits \cite{BrunoSommer2023}. For the moments this noise is small, but for the null-space specification tests of Sec.~\ref{sec:blue_results} it is not, so there we also use a nested jackknife. For each deleted configuration $i$, the central fit to the remaining $N-1$ configurations is redone with a temporal covariance re-estimated from those configurations alone, using their $N-1$ inner delete-one replicas (each of which retains $N-2$ configurations); the resulting $\theta^{(\rm n)}_i$ enter Eq.~(\ref{eq:jkcov}) in place of $\theta_i$. We denote the field-space covariances of the four responses from the two procedures by $\Sigma_{\rm fix}$ (temporal covariance held fixed, the conventional choice) and $\Sigma_{\rm nest}$ (nested jackknife, noise-inclusive). The cost is $N$ covariance estimates and refits per field. The deletion is common to all fields, so cross-field correlations are preserved. No regularization is applied: the 9- and 10-slice temporal covariances are full rank, and the correlated fits use eigen-decomposed whitening. We calibrate the resulting specification tests both against $\chi^2$ references and with full-pipeline configuration bootstraps (Sec.~\ref{sec:blue_results}). Section~\ref{sec:tempreg} repeats the analysis with uncorrelated and shrunk temporal covariances. For the learning curve, we draw 30 reproducible random subsets at each $N=100,200,400,600,800$ (random seed 20261008), re-extract the field responses using the same fit prescription, and optimize the O4-constrained weights from the subset covariance. The complete $N=992$ proton and $N=999$ neutron solutions serve as references rather than independent random subsets. The displayed subset bands span the 10th to 90th percentiles across the 30 draws; they are not confidence intervals for an asymptotic population optimum.

\section{Fixed field-space hierarchy}
\label{sec:fixedresults}

Table~\ref{tab:highhierarchy} summarizes the O2--O8 hierarchy after the normalization in Eq.~(\ref{eq:magresponse}). The estimators are nearly coincident relative to their marginal errors, but this visual agreement is not a specification test because the successive estimators are extremely correlated. The errors are delete-one jackknife errors with the temporal covariance held fixed; re-estimating it in every replica (nested jackknife) raises them by 2--3\% for the proton and by 3--11\% for the neutron, and full-pipeline bootstraps give spreads 0.93--1.00 times the nested errors (Sec.~\ref{sec:blue_results}).

\begin{table*}[t]
\caption{Magnetic-moment hierarchy in nuclear magnetons using Eq.~(\ref{eq:magresponse}), production windows. Parentheses give delete-one jackknife errors (full nonlinear refits in every replica, temporal covariance held fixed). The O4--O8 values differ by less than $5\times10^{-4}\,\mu_N$; the O2 values differ from them by 0.1--0.5\%.}
\label{tab:highhierarchy}
\centering
\begin{tabular}{rcccc@{\hspace{1.5em}}cccc}
\toprule
$m_\pi$ [MeV]&\multicolumn{4}{c}{Proton}&\multicolumn{4}{c}{Neutron}\\
\cmidrule(lr){2-5}\cmidrule(lr){6-9}
& O2 & O4 & O6 & O8 & O2 & O4 & O6 & O8\\
\midrule
338&1.959(41)&1.961(41)&1.961(41)&1.961(41)&-1.288(89)&-1.294(90)&-1.294(90)&-1.295(90)\\
362&1.937(38)&1.940(38)&1.940(38)&1.940(38)&-1.267(80)&-1.272(80)&-1.272(80)&-1.273(80)\\
384&1.917(35)&1.920(35)&1.920(35)&1.920(35)&-1.247(72)&-1.253(73)&-1.253(73)&-1.253(73)\\
405&1.897(32)&1.900(32)&1.900(32)&1.900(32)&-1.230(66)&-1.235(66)&-1.235(66)&-1.235(66)\\
444&1.861(29)&1.863(29)&1.863(29)&1.863(29)&-1.199(56)&-1.204(56)&-1.204(56)&-1.204(56)\\
693&1.624(14)&1.626(14)&1.626(14)&1.626(14)&-1.027(23)&-1.030(23)&-1.030(23)&-1.030(23)\\
\bottomrule
\end{tabular}
\end{table*}

The lightest-ensemble values are not intended as physical-point predictions. The underlying quenched, finite-volume, finite-spacing, heavier-than-physical-pion calculation has not been extrapolated to the continuum or physical quark masses. The purpose is to compare estimators on identical archival correlators. As a convention-independent cross-check, the lightest O4 ratio is $\mu_n/\mu_p=-1.294/1.961=-0.660$, compared with the experimental ratio $-1.913/2.793\simeq-0.685$; the two moments use different fit windows and slightly different configuration sets (999 and 992), and the comparison is descriptive only.

\subsection{Euclidean fit-window robustness}
\label{sec:windows}

The production windows, $t=12$--20 for the proton and $t=17$--26 for the neutron, are those of the original analysis and were fixed before any of the field-space diagnostics in this paper were computed. Table~\ref{tab:windows} repeats the complete analysis, with the same delete-one pipeline, while moving $t_{\min}$ at fixed $t_{\max}$. Differences from the production window are evaluated with paired jackknife replicas, which share configurations and therefore resolve much smaller shifts than the marginal errors suggest.

For the five lighter masses, moving $t_{\min}$ later by up to four slices changes the O6 moments by at most $1.9$ (proton) and $1.6$ (neutron) paired standard errors; moving it earlier, the proton window $t_{\min}=10$ is clearly excluded ($+2.9$ to $+4.1\sigma$), while $t_{\min}=11$ is not. The heaviest proton is not on a plateau: its O6 moment falls monotonically from 1.681 at $t_{\min}=10$ to 1.580 at $t_{\min}=15$ (1.588 at 16), with paired shifts of $-1.9$ to $-2.7\sigma$ at $t_{\min}=13$--15, so field-dependent excited-state contamination may contribute there; two-state fits (Sec.~\ref{sec:tempreg}) confirm that it does. Because this ensemble enters the medians and calibrations below, we state where excluding it changes them. The values for all six masses are in the reproducibility package. The O6 null-space statistic $\chi^2_{\rm O6}$ is window dependent with the temporal covariance held fixed (lightest proton 1.6--9.9, heaviest neutron 12.7--17.3). With the noise of the estimated covariance included it stays below the 5\% critical value of 3.84 in all but three of the 28 tabulated cases: the earliest heaviest-proton window ($\chi^2_{\rm O6}=5.9$ at $t=10$--20), the earliest lightest-neutron window (4.7 at $t=14$--26) and one heaviest-neutron window (3.9 at $t=15$--26). This agrees with Sec.~\ref{sec:blue_results}: departures from the O6 family are not resolved at the production windows, and there is marginal tension in only three windows, two of them the earliest.

\begin{table}[t]
\caption{Fit-window dependence of the O6 magnetic moment (nuclear magnetons) at the lightest and heaviest masses. Parentheses give marginal jackknife errors in units of the last digit; square brackets give the paired difference from the production window (rows without brackets) in units of its jackknife error. $\chi^2_{\rm O6}$ is the O6 null-space statistic (one degree of freedom), with the temporal covariance held fixed / with its estimation noise included.}
\label{tab:windows}
\centering
\begin{adjustbox}{max width=\columnwidth}
\begin{tabular}{lcccc}
\toprule
Window & O6, $m_\pi=338$ & O6, $m_\pi=693$ & $\chi^2_{{\rm O6},338}$ & $\chi^2_{{\rm O6},693}$\\
\midrule
\multicolumn{5}{l}{Proton}\\
10--20 & 2.050(27) [$+2.9$] & 1.681(11) [$+6.2$] & 9.9 / 2.9 & 17.2 / 5.9\\
11--20 & 1.981(33) [$+0.8$] & 1.648(12) [$+3.2$] & 4.6 / 1.2 & 5.1 / 1.7\\
12--20 & 1.961(41) & 1.626(14) & 4.0 / 1.0 & 0.6 / 0.2\\
13--20 & 1.945(50) [$-0.5$] & 1.608(17) [$-1.9$] & 3.8 / 1.0 & 0.1 / 0.0\\
14--20 & 1.905(61) [$-1.2$] & 1.588(20) [$-2.7$] & 3.2 / 0.9 & 3.9 / 0.9\\
15--20 & 1.959(79) [$0.0$] & 1.580(23) [$-2.5$] & 2.4 / 0.6 & 6.7 / 1.9\\
16--20 & 2.042(102) [$+0.9$] & 1.588(28) [$-1.6$] & 1.6 / 0.5 & 8.5 / 2.2\\
\multicolumn{5}{l}{Neutron}\\
14--26 & $-1.187(45)$ [$+1.4$] & $-1.019(14)$ [$+0.6$] & 13.0 / 4.7 & 15.9 / 3.5\\
15--26 & $-1.215(55)$ [$+1.1$] & $-1.013(16)$ [$+1.1$] & 11.5 / 3.8 & 17.3 / 3.9\\
16--26 & $-1.328(70)$ [$-0.6$] & $-1.026(19)$ [$+0.3$] & 7.3 / 2.1 & 15.1 / 3.4\\
17--26 & $-1.294(90)$ & $-1.030(23)$ & 6.6 / 2.1 & 13.1 / 2.9\\
18--26 & $-1.305(112)$ [$-0.2$] & $-1.037(27)$ [$-0.5$] & 5.7 / 2.0 & 12.7 / 3.0\\
19--26 & $-1.163(152)$ [$+1.1$] & $-1.019(34)$ [$+0.5$] & 4.9 / 1.5 & 13.0 / 3.2\\
20--26 & $-1.061(170)$ [$+1.6$] & $-1.004(38)$ [$+0.9$] & 2.6 / 1.0 & 13.7 / 3.3\\
\bottomrule
\end{tabular}
\end{adjustbox}
\end{table}

\section{Specification tests and covariance noise}
\label{sec:blue_results}

The fixed stencils agree with one another, but on archival data the target coefficient is unknown, so their agreement does not show that any of them is correct. We therefore test the response models directly, with statistics that do not depend on any optimized weight choice. For a response model with design matrix $A$ (the columns of Sec.~\ref{sec:constrained}), let $Q$ be any fixed basis for the left null space, $Q^TA=0$. If the model is correctly specified, the projected response $Q^T\bm y$ vanishes in expectation. With the aligned jackknife covariance $\Sigma$ of the four responses, we use
\begin{equation}
\chi^2_{\rm null}=(Q^T\bm y)^T(Q^T\Sigma Q)^{-1}(Q^T\bm y).
\label{eq:nullchisq}
\end{equation}
For O4, $A=(\bm 1,\bm x)$, so the null space has dimension two and the test has two degrees of freedom; for O6, $A=(\bm 1,\bm x,\bm x^2)$, leaving the single null direction $\bm n_6=(-64,84,-21,1)^T$ and one degree of freedom. The null bases depend only on the field geometry and the assumed response family, and the statistic does not depend on the choice of basis.

The null-space statistic also bounds discrepancies among estimators in the same constraint family. If $\bm w$ and $\bm v$ both satisfy $A^T\bm w=A^T\bm v=\bm b$, their difference $\bm d=\bm w-\bm v$ lies in the null space, and the Cauchy--Schwarz inequality in the null-space covariance metric gives, for $z=\bm d^T\bm y/\sqrt{\bm d^T\Sigma\bm d}$,
\begin{equation}
|z|\leq\sqrt{\chi^2_{\rm null}},
\label{eq:null_z_bound}
\end{equation}
with equality when the null space is one-dimensional. The bound holds when $z$ and $\chi^2_{\rm null}$ use the same $\Sigma$ and fixed weights. For O6, therefore, the difference between fixed O6 and any other O6-constrained estimator is the null test itself, not an independent diagnostic.

\begin{table*}[t]
\caption{Weight-free null-space specification tests for the nucleon response. The O4 model has two null degrees of freedom; O6 has one. Each entry gives the value with the temporal covariance held fixed in the jackknife, followed by the value with its estimation noise included (nested jackknife); $p$ values use the $\chi^2_2$ and $\chi^2_1$ references. Small $p$ values diagnose failure of the assumed truncated response family for the complete extraction pipeline, not uniquely a physical higher-order magnetic coefficient.}
\label{tab:nulltests}
\centering
\resizebox{\textwidth}{!}{%
\begin{tabular}{rcccc@{\hspace{1.5em}}cccc}
\toprule
$m_\pi$ [MeV]
& \multicolumn{4}{c}{Proton}
& \multicolumn{4}{c}{Neutron}\\
\cmidrule(lr){2-5}\cmidrule(lr){6-9}
& $\chi^2_{\rm O4}$ & $p_{\rm O4}$
& $\chi^2_{\rm O6}$ & $p_{\rm O6}$
& $\chi^2_{\rm O4}$ & $p_{\rm O4}$
& $\chi^2_{\rm O6}$ & $p_{\rm O6}$\\
\midrule
338 & 37.3 / 23.3 & $8.0\times10^{-9}$ / $8.8\times10^{-6}$ & 4.04 / 1.04 & 0.044 / 0.31 & 12.2 / 5.9 & 0.0022 / 0.053 & 6.61 / 2.07 & 0.010 / 0.15\\
362 & 49.8 / 37.0 & $<10^{-10}$ / $9.3\times10^{-9}$ & 3.85 / 0.98 & 0.050 / 0.32 & 16.3 / 8.4 & $2.9\times10^{-4}$ / 0.015 & 6.54 / 1.99 & 0.011 / 0.16\\
384 & 63.6 / 50.8 & $<10^{-10}$ / $<10^{-10}$ & 3.57 / 0.90 & 0.059 / 0.34 & 21.1 / 11.4 & $2.6\times10^{-5}$ / 0.0033 & 6.53 / 1.92 & 0.011 / 0.17\\
405 & 78.7 / 64.4 & $<10^{-10}$ / $<10^{-10}$ & 3.26 / 0.82 & 0.071 / 0.36 & 26.6 / 14.8 & $1.7\times10^{-6}$ / $6.2\times10^{-4}$ & 6.60 / 1.87 & 0.010 / 0.17\\
444 & 112.7 / 91.3 & $<10^{-10}$ / $<10^{-10}$ & 2.66 / 0.68 & 0.103 / 0.41 & 38.7 / 22.3 & $3.9\times10^{-9}$ / $1.4\times10^{-5}$ & 6.99 / 1.86 & 0.0082 / 0.17\\
693 & 618.8 / 252.0 & $<10^{-10}$ / $<10^{-10}$ & 0.64 / 0.21 & 0.42 / 0.65 & 165.8 / 130.1 & $<10^{-10}$ / $<10^{-10}$ & 13.12 / 2.94 & $2.9\times10^{-4}$ / 0.087\\
\bottomrule
\end{tabular}%
}
\end{table*}

With the temporal covariance held fixed, as in the production jackknife, O4 is rejected for every ensemble and O6 for every neutron ensemble ($p=2.9\times10^{-4}$--0.011), with the two lightest proton O6 tests near the five-percent boundary (Table~\ref{tab:nulltests}). These statistics omit the noise of the estimated temporal covariance. The moments are insensitive to it, but the null contrasts, being large-weight, nearly cancelling combinations of the four responses, are not: the error of the O6 contrast grows by a factor of 1.7--2.1, and the O4 statistics fall by factors of 1.2--2.5. With that noise included, O4 remains rejected for every ensemble except the lightest neutron ($p=0.053$), whereas O6 is rejected for none ($p=0.31$--0.65 for the proton, 0.09--0.17 for the neutron). With this pipeline the data therefore resolve departures from the O4 response family but not from the O6 family. Failure to reject is not evidence that O6 is correct: its single null direction cannot detect departures within the retained model space. The tests also assess the complete pipeline---correlated time fits, fit windows and field response---so the O4 rejections do not uniquely identify a physical $B^4$ magnetic term; field-dependent excited-state, boundary and other effects can project into the same null directions. Section~\ref{sec:tempreg} shows that the outcome of the O6 test depends on the temporal fit prescription.

We validated the nested correction with full-pipeline configuration bootstraps for the lightest and heaviest masses of both channels: each of 500 replicates resamples the configurations with replacement and reruns the complete analysis, including the temporal covariance. Table~\ref{tab:bootcal} compares the calibrations of the O6 test. The bootstrap spread of the O6 contrast is 2.2--2.4 times its fixed-covariance error and exceeds the nested-jackknife error by 5--34\%; the studentized bootstrap $p$ is lower than the nested one only for the heaviest neutron, which is the closest case ($p=0.054$--0.10). The studentized $p$ values change by at most 0.05 between 100 and 500 replicates. The O4 conclusions are unchanged: with the bootstrap covariance of the two O4 contrasts, O4 is rejected for the lightest proton ($p=7\times10^{-5}$) and at both heaviest masses ($p<10^{-28}$), and not for the lightest neutron ($p=0.14$). For the O6 moments themselves the bootstrap spread is 0.93--1.00 times the nested-jackknife errors (0.94--1.03 times the fixed-covariance ones), so the nested inflation of the moment errors (up to 11\% for the lightest neutron) is, if anything, conservative. For the null contrasts the asymmetry runs the other way: the nested jackknife understates their error by up to a third relative to the bootstrap. This only strengthens the non-rejection of O6, but it means that $\Sigma_{\rm nest}$, which we use below for the optimized weights, the frontier and the synthetic population, still slightly understates the noise in its smallest directions. For the eight ensembles without a bootstrap, the $\chi^2$ probabilities in Table~\ref{tab:nulltests} are asymptotic calibrations, supported by these four but not individually validated.

\begin{table}[t]
\caption{Calibration of the O6 null test for four ensembles: the $\chi^2_1$ tail probability with the temporal covariance held fixed, with the nested jackknife, the studentized full-pipeline bootstrap tail probability (500 replicates), and the $\chi^2_1$ probability using the bootstrap variance of the contrast.}
\label{tab:bootcal}
\centering
\begin{adjustbox}{max width=\columnwidth}
\setlength{\tabcolsep}{5pt}\begin{tabular}{lcccc}
\toprule
Ensemble & Fixed & Nested & Bootstrap & Bootstrap var.\\
\midrule
Proton, 338 MeV & 0.044 & 0.31 & 0.36 & 0.40\\
Proton, 693 MeV & 0.42 & 0.65 & 0.76 & 0.73\\
Neutron, 338 MeV & 0.010 & 0.15 & 0.15 & 0.24\\
Neutron, 693 MeV & $3\times10^{-4}$ & 0.087 & 0.054 & 0.10\\
\bottomrule
\end{tabular}
\end{adjustbox}
\end{table}

The field-space covariances are strongly hierarchical. For the lightest proton the eigenvalues relative to the largest are $1.9\times10^{-2}$, $1.3\times10^{-4}$ and $1.2\times10^{-6}$ with the temporal covariance fixed, and $2.3\times10^{-2}$, $2.9\times10^{-4}$ and $2.7\times10^{-6}$ with its noise included; the smallest varies by a factor of 2--3 under bootstrap resampling (10th--90th percentiles $1.3$--$3.1\times10^{-6}$). The limitation is the statistical uncertainty of these smallest-eigenvalue directions, which the large-weight contrasts and optimized estimators probe, not numerical inversion. Because the six valence masses share their gauge configurations, the tests across masses are correlated; we use their mass dependence to characterize the persistence of the model departure rather than combining the $p$ values into a global significance.

As a further diagnostic, we ask whether the O4 failure is already resolved among the three weaker fields. Restricting the analysis to $\{B,2B,4B\}$ leaves the single null direction $(4,-5,1)^T$ and the one-degree-of-freedom statistic for $4y(B)-5y(2B)+y(4B)$. With the covariance noise included, it gives $p=0.65$--0.93 for the five lighter proton ensembles and $p=0.063$--0.11 for the five lighter neutron ensembles (0.38--0.87 and 0.0005--0.0047 with the covariance fixed), so the four-field O4 rejection at the lighter masses comes from the strongest field. At the heaviest mass O4 is rejected even without $8B$, with the noise-inclusive covariance: $\chi^2_1=12.9$ ($p=3.4\times10^{-4}$) for the proton and 12.2 ($p=4.7\times10^{-4}$) for the neutron. The localization of the lighter-mass departure to $8B$ is not explained by the magnetic length alone, since $\ell_B$ is already comparable to or below the source-to-boundary distance at $2B$ and $4B$, where the three-field test passes (Sec.~\ref{sec:methods}). It is consistent with boundary or finite-volume effects that grow with the field, with higher-order response, or with field-dependent excited states, and does not identify the mechanism.

\subsection{Dependence on the temporal fit}
\label{sec:tempreg}

The noise that dominates the null contrasts enters through the correlated temporal fits, so we repeated the analysis with the temporal covariance of the double ratio replaced by its diagonal (uncorrelated fits) and shrunk halfway toward its diagonal \cite{Michael1994,MichaelMcKerrell1995,LedoitWolf2004}, in each case with both the fixed-covariance and the nested jackknife. The O6 moments move by up to two production standard errors (lightest proton 1.948(56) uncorrelated against 1.961(41); heaviest proton 1.598(18) against 1.626(14)), and their errors grow by 21--48\%. With the regularized covariances the nested jackknife changes the moment errors by at most 2\%, and the error of the O6 null contrast by factors of 0.8--1.5 for the proton and 1.0--2.1 for the neutron. O4 remains rejected for every ensemble. The O6 model, which the correlated fits do not reject, is now rejected at the heavier masses: with the noise included, uncorrelated fits give $\chi^2_{\rm O6}=4.4$--176 ($p\le0.036$) at the five heavier proton masses and 4.4--97 at the three heaviest neutron masses, but 2.3--3.7 ($p=0.06$--0.13) at the lightest proton and the three lightest neutron masses; half-shrunk fits reject only at the three heaviest proton masses and the heaviest neutron mass. At the heaviest mass the rejection is overwhelming in every variant ($\chi^2_{\rm O6}=52$--176) and insensitive to the covariance noise.

Departures from the O6 family are therefore resolved by these fits at the heavier masses, and not by the correlated production fits. We have not isolated why. The regularized fits weight the time slices differently, and for the heaviest proton, which is not on a plateau (Sec.~\ref{sec:windows}), field-dependent excited-state contamination is a plausible cause, which we test next. The null tests characterize the pipeline as a whole: their outcome is a property of the time fit as well as of the field response, and fixed-order truncation in $B$ is adequate only relative to a stated temporal analysis, whose own systematics are here as large as the effect being tested.

\paragraph{Two-state fits.} To test the role of excited states, we refit every double ratio as $R(t,B)=A\,e^{-m(B)t}[1+b(B)\,e^{-\Delta(B)t}]$ from $t_{\min}=8$ (and, as a check, 6) to the end of the production window, with the nested jackknife. The gap is refit in every replica under a Gaussian prior of width 30\% about a central value per ensemble, the median over fields of free-gap fits at $t_{\min}=6$; holding it fixed instead is a further variant. The fits describe the data ($\chi^2/{\rm dof}=0.8$--1.3 for the proton, 1.3--1.7 for the neutron) and give $a\Delta\simeq0.2$--0.35, larger at $t_{\min}=8$ than at 6, so more than one excited state contributes: these fits are a systematic check, not a replacement production analysis. Table~\ref{tab:twostate} compares them with the production fits. First, the heaviest mass ceases to be exceptional. Its $\chi^2_{\rm O4}$ falls from 252 (proton) and 130 (neutron) to 26 and 12, comparable to the lighter masses (12--23), so its outlying rejection in the production fits is mostly excited-state contamination. O4 remains rejected for every ensemble ($p\le0.003$; $p\le0.013$ in every variant) and O6 for none ($p=0.05$--0.99). With $t_{\min}=6$ the O6 test is marginal (lowest $p=0.004$, heaviest proton), and with one gap shared by the four fields it rejects, because the fitted gaps differ between fields: for the neutron the gap at $8B$ is 10--45\% below that at $B$, direct evidence of field-dependent excited states. Second, the magnitudes of the moments fall, by 4.3--4.8\% for the proton (2.0--5.3 production standard errors) and 3--10\% for the neutron; at the lightest mass $\mu_n/\mu_p$ moves from $-0.660$ to $-0.620$. The absolute moments therefore carry an excited-state systematic larger than their statistical errors. Third, the precision available to covariance optimization depends on the temporal analysis. With the two-state fits the O4-constrained optimum gains 6--30\% (proton) and 3--25\% (neutron), rather than 1--2\% (7--8\% at the heaviest mass), and the O6-constrained optimum 6--21\% for the proton and at most 1\% for the neutron; with $t_{\min}=6$ the O4 gain reaches 15--60\%. The refit gap fluctuates coherently across the four fields and the optimized weights partly cancel that common fluctuation, while moving the moments by 0.2--2.2 marginal standard errors from fixed O4. The estimator comparisons in the rest of the paper use the production fits and are conditional on them.

\begin{table*}[t]
\caption{Production single-exponential fits (1-st) against two-state fits (2-st; $t_{\min}=8$, gap refit under a 30\% prior), both with the nested jackknife: O6 moment in $\mu_N$, null-space statistics $\chi^2_{\rm O4}$ (2 degrees of freedom) and $\chi^2_{\rm O6}$ (1), and the precision gain in \% of the O4-constrained optimum over fixed O4. Ensembles are labelled by $m_\pi$ in MeV.}
\label{tab:twostate}
\centering
\setlength{\tabcolsep}{6pt}\begin{tabular}{lcccccccc}
\toprule
 & \multicolumn{2}{c}{O6 moment} & \multicolumn{2}{c}{$\chi^2_{\rm O4}$} & \multicolumn{2}{c}{$\chi^2_{\rm O6}$} & \multicolumn{2}{c}{Gain (\%)}\\
Ensemble & 1-st & 2-st & 1-st & 2-st & 1-st & 2-st & 1-st & 2-st\\
\midrule
Proton, 338 & 1.961(42) & 1.877(48) & 23 & 12 & 1.04 & 0.05 & 1.7 & 6\\
Proton, 362 & 1.940(38) & 1.855(46) & 37 & 17 & 0.98 & 0.01 & 1.4 & 9\\
Proton, 384 & 1.920(36) & 1.834(45) & 51 & 20 & 0.90 & 0.00 & 1.3 & 12\\
Proton, 405 & 1.900(33) & 1.814(44) & 64 & 21 & 0.82 & 0.02 & 1.4 & 15\\
Proton, 444 & 1.863(29) & 1.775(42) & 91 & 23 & 0.68 & 0.10 & 1.7 & 20\\
Proton, 693 & 1.626(15) & 1.548(28) & 252 & 26 & 0.21 & 0.15 & 6.7 & 30\\
\midrule
Neutron, 338 & $-1.294(100)$ & $-1.163(34)$ & 6 & 19 & 2.07 & 3.81 & 1.4 & 3\\
Neutron, 362 & $-1.272(89)$ & $-1.157(32)$ & 8 & 21 & 1.99 & 3.49 & 1.2 & 4\\
Neutron, 384 & $-1.253(80)$ & $-1.150(30)$ & 11 & 22 & 1.92 & 3.15 & 1.3 & 6\\
Neutron, 405 & $-1.235(73)$ & $-1.142(29)$ & 15 & 22 & 1.87 & 2.82 & 1.5 & 8\\
Neutron, 444 & $-1.204(61)$ & $-1.126(26)$ & 22 & 21 & 1.86 & 2.32 & 2.3 & 12\\
Neutron, 693 & $-1.030(24)$ & $-0.997(16)$ & 130 & 12 & 2.94 & 0.08 & 7.6 & 25\\
\bottomrule
\end{tabular}
\end{table*}

\section{Covariance-optimized estimators}
\label{sec:blue}

\subsection{Weights and leverage}

With four fields and the O4 constraints, two weight degrees of freedom remain, and the minimum-variance (BLUE) solution, Eq.~(\ref{eq:blueweights}), is the correlated linear fit in $B^2$ to all four fields. For the lightest proton, with the conventional field-space covariance from the fixed-temporal-covariance jackknife,
\begin{equation}
\bm w_{\rm BLUE}\simeq(56.37,-82.74,30.77,-3.40),
\end{equation}
with $L_4\simeq-7.3\times10^3$ and $L_6\simeq-7.7\times10^5$, against $L_4=-4$ and $L_6=-20$ for the fixed O4 stencil; the lightest-neutron solution is $(14.85,-30.04,19.39,-3.203)$ with $L_4\simeq-8.6\times10^3$ and $L_6\simeq-7.6\times10^5$. With the noise-inclusive covariance the weights are smaller, for example $(24.7,-37.6,15.9,-2.0)$ with $L_6\simeq-4.7\times10^5$ for the lightest proton and $(12.8,-17.5,6.4,-0.71)$ with $L_6\simeq-1.6\times10^5$ for the lightest neutron, but still far from O4. Full-precision weights are included in the reproducibility package. Negative BLUE weights can be legitimate consequences of strongly correlated observations \cite{ValassiChierici2014}; they do not depend on the coordinates used for the optimization.

Throughout, a displacement means BLUE minus fixed. The differences obey Eq.~(\ref{eq:null_z_bound}). With $\Sigma_{\rm fix}$ the O4 differences are $-4.2$ to $-8.4\sigma$ for the proton and $+3.4$ to $+12.7\sigma$ for the neutron; with $\Sigma_{\rm nest}$ they are $-4.8$ to $-15.8\sigma$ and $-0.3$ to $+10.9\sigma$, each below the corresponding $\sqrt{\chi^2_{\rm O4}}$ in magnitude. For O6 their magnitudes equal $\sqrt{\chi^2_{\rm O6}}$: 0.8--2.0$\sigma$ (proton) and 2.6--3.6$\sigma$ (neutron) with $\Sigma_{\rm fix}$, 0.5--1.7$\sigma$ with $\Sigma_{\rm nest}$. These $\sigma$ are paired errors of the difference, small because the two estimators are almost perfectly correlated. Our jackknife differences satisfy the Hausman identity \cite{Hausman1978}, ${\rm Var}(\widehat\theta_{\rm fixed}-\widehat\theta_{\rm BLUE})={\rm Var}(\widehat\theta_{\rm fixed})-{\rm Var}(\widehat\theta_{\rm BLUE})$, which holds for weights held fixed and evaluated with the covariance that defines them; for weights reoptimized on each sample it is not exact, so a BLUE error $\rho\,\sigma_{\rm O4}$ implies a paired error $\sigma_{\rm O4}\sqrt{1-\rho^2}$; a 1.5\% gain thus turns $7\sigma$ paired into about $1.2\sigma$ marginal. Table~\ref{tab:displacement} gives the displacements in nuclear magnetons and in units of the marginal fixed-O4 error. With $\Sigma_{\rm nest}$, the BLUE-O4 proton moments lie 0.037--0.052~$\mu_N$ (1.9--2.8\%, 0.9--1.8 marginal standard errors) below fixed O4 at the five lighter masses and 0.083~$\mu_N$ (5.7) at the heaviest; the neutron moments differ by at most 0.046~$\mu_N$ (0.8) at the lighter masses. With $\Sigma_{\rm fix}$ the neutron moments are smaller in magnitude by 0.15--0.17~$\mu_N$, 11--15\% or 1.6--6.8 marginal standard errors. Equation~(\ref{eq:leverageidentity}) reproduces every BLUE-O4 minus O8 difference to numerical precision.

\begin{table*}[t]
\caption{Displacement of the unrestricted O4-constrained BLUE moment from fixed O4 (BLUE minus fixed), in nuclear magnetons, for both field-space covariances. Square brackets give the displacement in units of the marginal fixed-O4 jackknife error and, after the semicolon, of the paired error of the difference. Weights held fixed in the jackknife.}
\label{tab:displacement}
\centering
\begin{tabular}{rcccc}
\toprule
& \multicolumn{2}{c}{Proton} & \multicolumn{2}{c}{Neutron}\\
\cmidrule(lr){2-3}\cmidrule(lr){4-5}
$m_\pi$ [MeV] & $\Sigma_{\rm fix}$ & $\Sigma_{\rm nest}$ & $\Sigma_{\rm fix}$ & $\Sigma_{\rm nest}$\\
\midrule
338 & $-0.065$ [$-1.6$; $-6.1$] & $-0.037$ [$-0.9$; $-4.8$] & $+0.145$ [$+1.6$; $+3.4$] & $-0.004$ [$-0.0$; $-0.3$] \\
362 & $-0.070$ [$-1.9$; $-6.8$] & $-0.039$ [$-1.0$; $-6.1$] & $+0.153$ [$+1.9$; $+4.0$] & $+0.008$ [$+0.1$; $+0.6$] \\
384 & $-0.076$ [$-2.2$; $-7.4$] & $-0.041$ [$-1.2$; $-7.1$] & $+0.158$ [$+2.2$; $+4.6$] & $+0.019$ [$+0.2$; $+1.5$] \\
405 & $-0.081$ [$-2.5$; $-7.9$] & $-0.044$ [$-1.3$; $-8.0$] & $+0.162$ [$+2.4$; $+5.1$] & $+0.028$ [$+0.4$; $+2.3$] \\
444 & $-0.090$ [$-3.1$; $-8.4$] & $-0.052$ [$-1.8$; $-9.5$] & $+0.168$ [$+3.0$; $+6.2$] & $+0.046$ [$+0.8$; $+3.6$] \\
693 & $-0.035$ [$-2.5$; $-4.2$] & $-0.083$ [$-5.7$; $-15.8$] & $+0.158$ [$+6.8$; $+12.7$] & $+0.100$ [$+4.2$; $+10.9$] \\
\bottomrule
\end{tabular}
\end{table*}

\subsection{The leverage frontier}
\label{sec:tau}

\begin{figure*}[t]
\centering\includegraphics[width=0.85\textwidth]{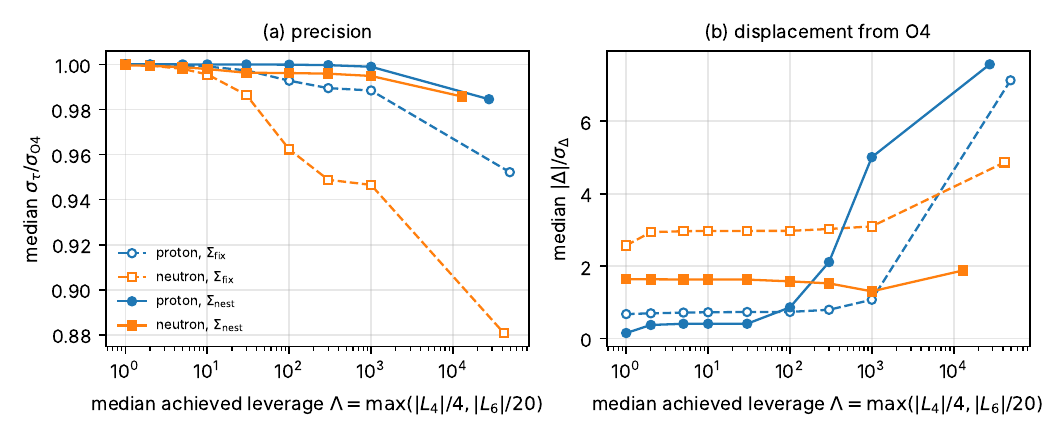}
\caption{The leverage frontier. (a) Median standard-error ratio relative to fixed O4 and (b) median magnitude of the displacement from fixed O4 in units of its paired error, against the median achieved leverage $\Lambda$, with $\Sigma_{\rm fix}$ (open symbols, dashed) and $\Sigma_{\rm nest}$ (filled, solid). The six masses within a channel share their gauge configurations and are not independent.}
\label{fig:tau}
\end{figure*}

To see how precision is traded for leverage, we minimize the estimated variance subject to the O4 constraints and
\begin{equation}
|L_4(\bm w)|\le4\tau,\qquad |L_6(\bm w)|\le20\tau,
\label{eq:tau}
\end{equation}
for $\tau$ from 1 to 1000 and without bounds, and record the achieved leverage
\begin{equation}
\Lambda=\max(|L_4|/4,\,|L_6|/20).
\end{equation}
 The bounds are diagnostic, not a prescription for choosing $\tau$. Table~\ref{tab:tau} and Fig.~\ref{fig:tau} give the results for both field-space covariances, with the optimized weights held fixed in the jackknife.

\begin{table}[t]
\caption{Median standard-error ratios $\sigma_\tau/\sigma_{\rm O4}$ over the six pion masses along the leverage frontier, with the conventional (fixed-temporal-covariance) and the noise-inclusive field-space covariance; weights held fixed in the jackknife.}
\label{tab:tau}
\centering
\begin{tabular}{rcccc}
\toprule
& \multicolumn{2}{c}{$\Sigma_{\rm fix}$ (conventional)} & \multicolumn{2}{c}{$\Sigma_{\rm nest}$ (noise-inclusive)}\\
\cmidrule(lr){2-3}\cmidrule(lr){4-5}
$\tau$&Proton&Neutron&Proton&Neutron\\\midrule
1&0.9998&1.0000&1.0000&0.9996\\
2&0.9997&0.9995&1.0000&0.9994\\
5&0.9995&0.9980&1.0000&0.9988\\
10&0.9990&0.9955&1.0000&0.9979\\
30&0.9973&0.9864&0.9999&0.9963\\
100&0.9927&0.9623&0.9999&0.9961\\
300&0.9894&0.9488&0.9996&0.9958\\
1000&0.9884&0.9465&0.9989&0.9948\\
Unrestricted&0.9522&0.8808&0.9845&0.9857\\
\bottomrule
\end{tabular}
\end{table}

With the conventional covariance, variance reduction develops gradually as leverage is admitted: at $\tau=100$ the median gains are 0.7\% (proton) and 3.8\% (neutron), at $\tau=1000$ 1.2\% and 5.3\%, and without bounds 4.8\% and 12\%, at leverages four orders of magnitude above fixed O4. With the noise-inclusive covariance almost no precision is available: the median gain is at most 0.1\% (proton) and 0.5\% (neutron) up to $\tau=1000$, and 1.5\% and 1.4\% without bounds (1--2\% at the five lighter masses, 7--8\% at the heaviest); without the heaviest mass, the unrestricted medians are 1.4\% for both. These gains refer to the production temporal fits; with two-state fits they are larger (Sec.~\ref{sec:tempreg}). The displacement from fixed O4 [Fig.~\ref{fig:tau}(b)] does not track the gain. For the neutron it is already 2.6$\sigma$ (conventional) or 1.6$\sigma$ (noise-inclusive) at $\tau=1$, where the gain is below 0.05\%, and stays nearly flat up to $\tau=1000$: any admissible weight other than O4 picks up the resolved O4 null direction, as Eq.~(\ref{eq:null_z_bound}) implies. For the proton it grows with leverage, to a median of $7.6\sigma$ (paired; $7.1\sigma$ without the heaviest mass) without bounds with $\Sigma_{\rm nest}$, for a 1.5\% gain, or about one marginal standard error (Table~\ref{tab:displacement}). If the unrestricted weights are re-derived in each jackknife replica, the median ratios become 1.18 and 0.98 (conventional) and 1.13 and 1.19 (noise-inclusive).

\subsection{Independent evaluation}

A delete-one replica is not independent of the sample that fixed the weights. We therefore split the configurations of the lightest proton and neutron at random into two halves, 20 times; the unrestricted O4-constrained weights are fitted on one half and applied unchanged to the other, and their error there is compared with fixed O4 on the same half, with temporal-covariance noise included. With weights from the conventional covariance, the median error ratio is 1.04 for the proton and 1.16 for the neutron (never below one for the neutron, and below one in 3 of 20 proton splits), while the same weights would appear 5--9\% more precise if judged on the evaluation half's own covariance. With weights from the noise-inclusive covariance there is no gain to lose: the median ratio is 1.00 for the proton and 0.99 for the neutron. The displacement from fixed O4 persists on independent data for the proton (median $-3.0\sigma$ with the conventional weights, $-3.2\sigma$ with the noise-inclusive ones); for the neutron it is $+1.4\sigma$ with the conventional weights and vanishes with the noise-inclusive ones. The variance reduction therefore does not generalize, while the projection onto the resolved O4 departure does.

\subsection{Dependence on ensemble size}
\label{sec:learning}

\begin{figure}[t]
\centering\includegraphics[width=\columnwidth]{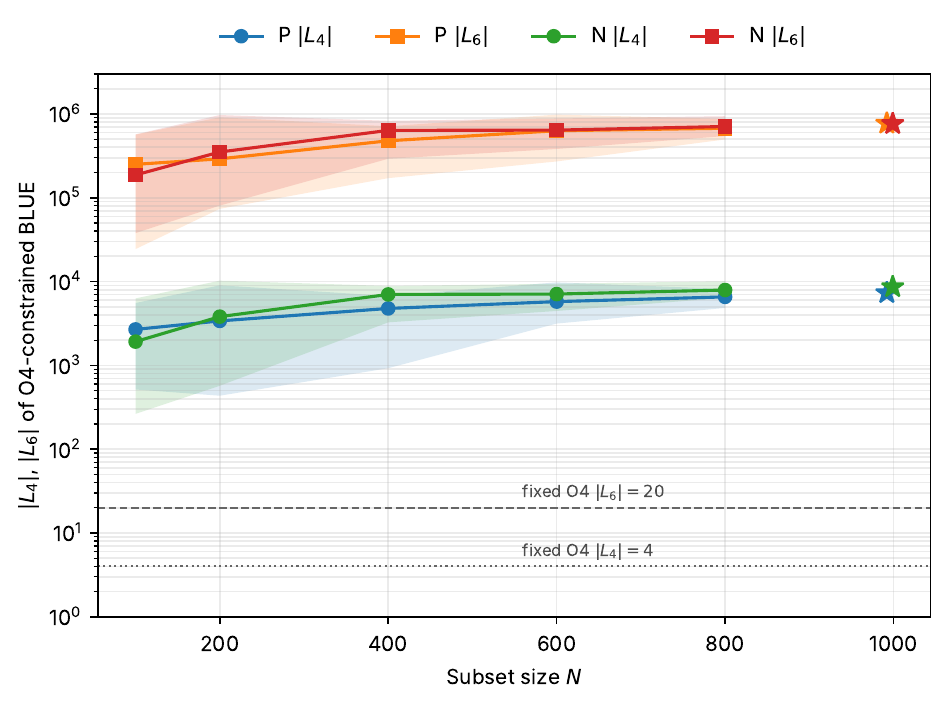}
\caption{Magnitude of the omitted-mode leverage of the O4-constrained BLUE weights against configuration subset size for the lightest proton and neutron (conventional covariance). Points are medians of $|L_4|$ and $|L_6|$ over 30 random subsets, bands their 10th--90th percentiles; stars mark the full samples. Horizontal lines mark the fixed O4 stencil, $|L_4|=4$ and $|L_6|=20$.}
\label{fig:learning}
\end{figure}

At subset sizes $N=100$--800 we repeat the complete extraction and optimize the O4-constrained weights from the subset covariance, with the conventional jackknife. For the lightest proton the median first two weights move from $(22,-31)$ at $N=100$ to $(46,-68)$ at $N=800$, against $(56,-83)$ for the full sample; for the lightest neutron from $(6,-7)$ to $(16,-30)$, against $(15,-30)$. The magnitude of the leverage grows with $N$ (Fig.~\ref{fig:learning}): larger samples do not drive the covariance optimum toward the low-leverage stencil. The simpler reading, which the synthetic study below supports, is that the empirical covariance increasingly resolves a high-leverage population optimum. The learning curve uses the conventional covariance; with the noise-inclusive one the population optimum is also high leverage (Sec.~\ref{sec:synthetic}). The 30 random subsets at each size overlap heavily and are not independent.

\subsection{Known-covariance synthetic study}
\label{sec:synthetic}

To separate estimator variance, covariance-estimation noise and omitted-mode displacement without relying on archival data whose true moments are unknown, we generate synthetic field-space ensembles with a known population covariance and known response. The population covariance is the noise-inclusive field-space covariance of the lightest proton or neutron; the truth is $y(x)=y_0+c_1x+c_2x^2+c_3x^3$, with coefficients from the exact cubic through the measured responses, and with $c_3$ as measured or tripled. Each mock ensemble draws $n$ configurations from a Gaussian with per-configuration covariance $n$ times the population covariance, so that the mock mean has the population covariance at every $n$ and only the covariance estimate degrades as $n$ decreases; estimators use the mock mean and the mock's estimated covariance of the mean. The mocks are drawn at the level of the four field responses, so they include the noise of the estimated $4\times4$ field-space covariance but not that of the temporal fits (Sec.~\ref{sec:blue_results}). We compare the fixed stencils, BLUE with the true and the estimated covariance, and correlated polynomial fits, using 1500--2000 mocks per setting (Monte Carlo error of the displacements about $0.03\sigma$).

Table~\ref{tab:synthetic} gives the displacement from the true $y_0$, in units of each estimator's standard deviation over the mocks, and the root-mean-square error relative to fixed O6 at $n=992$. The fixed stencils are undisplaced within the Monte Carlo error. BLUE-O4 has the same constraints as O4 but is displaced by up to $4.7\sigma$, with a root-mean-square error up to 4.8 times that of O6. This happens with the \emph{true} covariance, so it is omitted-mode leverage acting on a genuine $x^2$ and $x^3$ response, not covariance estimation; its size depends on the combination $c_2L_4+c_3L_6$, which can cancel (neutron, measured $c_3$). BLUE-O6 is within about $0.2\sigma$ of O6 except when the omitted term is enlarged. At $n=100$ the fixed-weight errors of the optimized estimators understate their actual spread by factors of up to six when the omitted terms are large, while BLUE-O6 remains reliable (reported/actual error 0.89--0.97).

\begin{table}[t]
\caption{Synthetic ensembles ($n=992$) with known covariance and response. For each truth, the first column is the mean displacement from the true $y_0$ in units of the estimator's standard deviation over the mocks (Monte Carlo error about 0.03), the second the root-mean-square error relative to fixed O6. ``Data'' uses the measured cubic coefficient; ``$3\times$'' triples it.}
\label{tab:synthetic}
\centering
\setlength{\tabcolsep}{3pt}\resizebox{\columnwidth}{!}{%
\begin{tabular}{lcccccccc}
\toprule
& \multicolumn{4}{c}{Proton} & \multicolumn{4}{c}{Neutron}\\
\cmidrule(lr){2-5}\cmidrule(lr){6-9}
& \multicolumn{2}{c}{Data} & \multicolumn{2}{c}{$3\times$} & \multicolumn{2}{c}{Data} & \multicolumn{2}{c}{$3\times$}\\
Estimator & disp. & RMSE & disp. & RMSE & disp. & RMSE & disp. & RMSE\\
\midrule
O4 & $-0.02$ & 1.00 & 0.00 & 1.00 & $-0.02$ & 1.00 & $-0.01$ & 1.00\\
O6 & $-0.02$ & 1.00 & 0.00 & 1.00 & $-0.02$ & 1.00 & $-0.01$ & 1.00\\
O8 & $-0.02$ & 1.00 & 0.00 & 1.00 & $-0.02$ & 1.00 & $-0.01$ & 1.00\\
BLUE-O4, true $\Sigma$ & $-0.92$ & 1.34 & $-4.73$ & 4.75 & $-0.06$ & 0.98 & $-3.90$ & 3.96\\
BLUE-O4, estimated $\Sigma$ & $-0.90$ & 1.35 & $-3.66$ & 4.82 & $-0.06$ & 0.99 & $-2.43$ & 4.17\\
BLUE-O6, estimated $\Sigma$ & $-0.04$ & 1.00 & $-0.04$ & 1.00 & $-0.22$ & 1.01 & $-0.62$ & 1.17\\
BLUE-1, estimated $\Sigma$ & $-5.54$ & 5.13 & 1.57 & 2.21 & 1.25 & 1.47 & 5.72 & 10.45\\
\bottomrule
\end{tabular}}
\end{table}

\paragraph{How much precision is available.} Even if the covariance estimated from these data were the exact population covariance, the optimum would gain little. For the four ensembles with full-pipeline bootstraps, the population BLUE-O4 has a standard error 0.92--0.99 times that of fixed O4, whether the population covariance is taken as the nested-jackknife or the bootstrap covariance (10th--90th percentiles over bootstrap resamplings of the latter: 0.92--0.997), while its leverage is $|L_6|\simeq(0.8$--$14)\times10^5$ in every case. The population BLUE-O6 gains at most 2\% over fixed O6. The high leverage is therefore not specific to one point estimate of the covariance, but neither is there much precision to buy with it.

\paragraph{Specification test as a pre-test.} The large displacements above occur in a response family that the data reject. To ask whether the null test detects a misspecification before the optimized estimator is displaced, we scaled the omitted coefficients ($c_2$ and $c_3$ together, or $c_3$ alone) by factors 0--3 and recorded, at $n=992$, the rejection rate of the O4 and O6 null tests at 5\% together with the displacements. In every scenario in which BLUE-O4 is displaced by more than $0.5\sigma$, the O4 test rejects in at least 99\% of the mocks, for both channels; at displacements of $0.4$--$0.5\sigma$ it rejects in 59--72\%. BLUE-O6 is displaced by at most $0.6\sigma$ even when the O6 test rejects in 99\% of the mocks. In this idealized setting, a specification test followed by optimization within the accepted family would keep the displacement small. On the real data the tests have less power, because the temporal-covariance noise roughly doubles the error of the null contrasts, and, as Sec.~\ref{sec:tempreg} shows, which family is accepted depends on the time fit.

\paragraph{Field design.} The leverages scale with the field design: with the geometric design $x=(1,4,16,64)$ the fixed O4 stencil has $L_6=-20$, and with equally spaced fields $(2,4,6,8)B_0$ it would have $-1280$. To compare designs, we write the measured covariance as that of a random cubic curve in $x$, $\Sigma=VKV^T$ with $V$ the $4\times4$ Vandermonde matrix, and evaluate it at other fields in the same range. In this common-curve model the population BLUE is the same linear functional of the response curve for any four distinct fields, so its displacement $c_2L_4+c_3L_6$, in absolute units, is independent of the design; what changes is the leverage and precision of the fixed stencils, which for $(2,4,6,8)B_0$ lose up to 1\% in precision. Because $\Sigma=VKV^T$ makes the field-space covariance equivalent across designs by construction, this is a property of the model rather than a test: it shows only that, for noise of this form, the geometric spacing does not produce the displacement. Testing more would require a design-sensitive noise model or data at a second design.

\section{Known-answer controls: the decuplet}
\label{sec:decuplet}

The tests above detect departures from a truncated response model, but on the nucleon the target is unknown, so they cannot show that any estimator returns the correct value. The decuplet correlators provide cases in which the answer is known exactly.

\subsection{Exact first-order identities}

The production code builds the $\Delta^{++}$ correlator from the contraction $F(S_u,S_u,\tilde S_u)$ and the $\Delta^+$ correlator as the symmetric average $\tfrac13[F(S_u,S_u,\tilde S_d)+F(S_d,S_u,\tilde S_u)+F(S_u,S_d,\tilde S_u)]$ over the position of the $d$ quark (Appendix~\ref{app:identity}); the $\Delta^0$ and $\Delta^-$ are built in the same way with two and three $d$ lines. With degenerate valence masses and the field coupled only to the valence propagators, $S_q=S+e_q\,\delta S+O(e_q^2)$ with the same $S$ and $\delta S$ for both flavors. Because each correlator is linear in the charge weights of its lines with a common kernel, its first-order field response is proportional to the mean charge of its quarks: $e_u$, $\tfrac13(2e_u+e_d)$, $\tfrac13(e_u+2e_d)$ and $e_d$. Hence, configuration by configuration and at every time slice,
\begin{equation}
\frac{\mu_{\Delta^+}}{\mu_{\Delta^{++}}}=\frac12,\qquad \frac{\mu_{\Delta^0}}{\mu_{\Delta^{++}}}=0,\qquad \frac{\mu_{\Delta^-}}{\mu_{\Delta^{++}}}=-\frac12
\label{eq:deltaratio}
\end{equation}
for the weak-field coefficients. Since the four correlators are identical at zero field, the relations carry over to the linear energy coefficient of each non-degenerate level, without assumptions about ground-state dominance. Isospin alone would allow $\mu=a+bI_3$; linearity in the quark charges with a common kernel fixes $b/a$. The relations hold because the field couples only to the valence quarks; with charged sea quarks a contribution common to all four states would break them. They are independent of quenching, volume, lattice spacing and the overall normalization of Eq.~(\ref{eq:magresponse}), and they hold for any spin projection of the interpolating fields. The interpolating field is the $z$ component of the Rarita--Schwinger field, $\chi_3=\epsilon^{abc}(q^{aT}C\gamma_3q^b)q^c$, without spin-$\tfrac32$ projection, and the code stores its positive-parity diagonal Dirac components $G_{11}$ and $G_{22}$ (Appendix~\ref{app:identity}); with the field along $z$ these have $J_z=+\tfrac12$ and $-\tfrac12$. We combine them in the double ratio, Eq.~(\ref{eq:doubleratio}), as for the nucleon. For a spin-$\tfrac32$ state the $J_z=\pm\tfrac12$ splitting is $\tfrac23\mu_\Delta B$ rather than $2\mu B$, so with the nucleon normalization, Eq.~(\ref{eq:magresponse}), the reduced response gives $\mu_\Delta/3$; we use only ratios, in which this factor cancels. On the correlators themselves, the configuration-level deviation of the field-odd ratio $[C_{\Delta^+}(B)-C_{\Delta^+}(-B)]/[C_{\Delta^{++}}(B)-C_{\Delta^{++}}(-B)]$ from $\tfrac12$ scales as $B^2$ (at $t=10$ and the lightest mass, successive field doublings multiply its median by 4.0, 4.0 and 4.1; by $t=18$ the factors grow to 4.1, 4.2 and 4.6) and vanishes as $B\to0$.

The identities constrain the first-order coefficients only. The $\Delta^+$ and $\Delta^0$ have nucleon partners with the same quantum numbers, 160--260~MeV lower on these ensembles (Sec.~\ref{sec:methods}), to which the field can mix them through the $N\to\Delta$ magnetic transition \cite{Batelaan2023,DeshmukhTiburzi2018}; the $\Delta^{++}$ and $\Delta^-$ have none. Because the nucleon moment shifts its two spin states in opposite directions, such mixing contributes a spin-odd term to the energy splitting at third order in $B$, and higher terms at higher order. We do not establish this mechanism independently, but it is consistent with what follows: at the lightest mass $y(B)$ for the $\Delta^+$ rises by 58\% between $B$ and $8B$ while that of the $\Delta^{++}$ changes by 3\% [Fig.~\ref{fig:decuplet}(a)], and the coefficient of the $B^2$ deviation above grows with Euclidean time. Earlier background-field calculations of decuplet moments include Refs.~\cite{Lee2005BaryonMom,Aubin2009Delta}.

\subsection{Fixed stencils}

We analyze each state with the double-ratio pipeline of Sec.~\ref{sec:methods}, with delete-one replicas aligned across states on their common configurations (892 for $\Delta^+$/$\Delta^{++}$, 871 for $\Delta^0$, 901 for $\Delta^-$) and windows $t=10$--18 and 12--20 chosen from the $\Delta$ plateau. Both states of a pair are fitted to the double ratio exactly as defined in Eq.~(\ref{eq:doubleratio}); because the $\Delta^+$ ratio is not a single exponential at the larger fields, fitting its inverse instead shifts the ratio test by several standard errors. Each state's correlated fit uses its own temporal covariance, but the configuration-level fluctuations of the two field-odd responses are proportional at first order, so the identity survives the fits at that order. The ratio is formed replica by replica, so its error reflects the strong correlation between the states: the marginal moment errors are 1--6\%, while the ratio is determined to $2.5\times10^{-5}$--$3.8\times10^{-4}$ for O4--O8. At that precision the temporal-covariance noise matters, so all errors in this section come from the nested jackknife; for the O4--O8 ratios it doubles the fixed-covariance errors.

\begin{table*}[t]
\caption{Deviation of the decuplet ratio $r=\mu_{\Delta^+}/\mu_{\Delta^{++}}$ from the exact value $\tfrac12$, in units of $10^{-4}$, with paired jackknife errors in parentheses (including the noise of the estimated temporal covariance) and the deviation in standard errors in square brackets. Three of the six masses are shown; the others are in the reproducibility package.}
\label{tab:decuplet}
\centering
\begin{tabular}{ccllll}
\toprule
Window & $m_\pi$ [MeV] & O2 & O4 & O6 & O8\\
\midrule
10--18 & 338 & 72.3(4.7) [15] & 0.4(1.1) [0.4] & $-1.10(94)$ [$-1.2$] & $-1.09(91)$ [$-1.2$]\\
       & 444 & 62.7(3.2) [20] & 0.71(66) [1.1] & $-0.60(54)$ [$-1.1$] & $-0.62(53)$ [$-1.2$]\\
       & 693 & 41.8(1.6) [25] & 0.28(31) [0.9] & $-0.40(26)$ [$-1.6$] & $-0.40(25)$ [$-1.6$]\\
12--20 & 338 & 124.0(12.4) [10] & 1.0(3.8) [0.3] & $-2.9(3.4)$ [$-0.9$] & $-3.0(3.3)$ [$-0.9$]\\
       & 444 & 101.8(7.3) [14] & 1.6(2.0) [0.8] & $-1.5(1.7)$ [$-0.9$] & $-1.6(1.7)$ [$-1.0$]\\
       & 693 & 63.1(3.3) [19] & 0.87(60) [1.4] & $-0.62(49)$ [$-1.3$] & $-0.68(48)$ [$-1.4$]\\
\bottomrule
\end{tabular}
\end{table*}
\begin{table*}[t]
\caption{Absolute control: $r_0=\mu_{\Delta^0}/\mu_{\Delta^{++}}$, exactly zero at first order, in units of $10^{-4}$, with nested-jackknife errors and deviations in standard errors, window 10--18. Fixed stencils and per-state covariance-optimized estimators (noise-inclusive covariance, fixed weights).}
\label{tab:delta0}
\centering
\begin{tabular}{rllllll}
\toprule
$m_\pi$ & O2 & O4 & O6 & O8 & BLUE-O4 & BLUE-O6\\
\midrule
338 & $-52.14(1.90)$ [$-$27.5] & $-0.96(0.07)$ [$-$13.8] & $0.001(0.019)$ [0.1] & $0.035(0.014)$ [2.6] & $-0.084(0.007)$ [$-$12.5] & $0.078(0.010)$ [7.7]\\
362 & $-50.23(1.75)$ [$-$28.8] & $-0.90(0.06)$ [$-$13.9] & $0.004(0.017)$ [0.2] & $0.034(0.012)$ [2.7] & $-0.078(0.006)$ [$-$12.8] & $0.073(0.009)$ [7.8]\\
384 & $-48.46(1.62)$ [$-$30.0] & $-0.84(0.06)$ [$-$14.1] & $0.006(0.016)$ [0.4] & $0.033(0.012)$ [2.8] & $-0.072(0.006)$ [$-$13.0] & $0.068(0.009)$ [8.0]\\
405 & $-46.81(1.51)$ [$-$31.1] & $-0.78(0.06)$ [$-$14.1] & $0.009(0.015)$ [0.6] & $0.032(0.011)$ [3.0] & $-0.067(0.005)$ [$-$13.3] & $0.063(0.008)$ [8.0]\\
444 & $-43.83(1.33)$ [$-$33.1] & $-0.68(0.05)$ [$-$14.2] & $0.012(0.013)$ [1.0] & $0.030(0.009)$ [3.2] & $-0.058(0.004)$ [$-$13.7] & $0.055(0.007)$ [8.1]\\
693 & $-28.60(0.70)$ [$-$40.8] & $-0.27(0.02)$ [$-$11.1] & $0.022(0.006)$ [3.7] & $0.021(0.004)$ [5.2] & $-0.021(0.001)$ [$-$15.0] & $0.019(0.002)$ [7.8]\\
\bottomrule
\end{tabular}
\end{table*}

\begin{figure*}[t]
\centering\includegraphics[width=\textwidth]{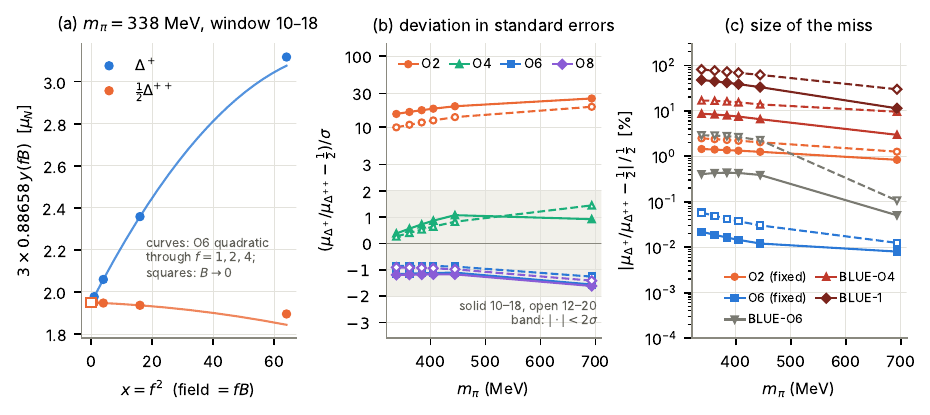}
\caption{(a) $3\times0.88658\,y(fB)$, which for the $J_z=\pm\tfrac12$ components of a spin-$\tfrac32$ state is the field-dependent moment in nuclear magnetons [Eq.~(\ref{eq:magresponse}) and Sec.~\ref{sec:decuplet}], against $x=f^2$ for the $\Delta^+$ and $\tfrac12\Delta^{++}$ at $m_\pi=338$~MeV, window 10--18. Curves: the O6 quadratic through $f=1,2,4$; squares: its $B\to0$ intercept, which coincides for the two states. (b) Deviation of $\mu_{\Delta^+}/\mu_{\Delta^{++}}$ from $\tfrac12$ in paired standard errors, including temporal-covariance noise, for each fixed stencil; solid symbols 10--18, open symbols 12--20; shading marks $|\cdot|<2\sigma$; symmetric logarithmic axis. (c) Size of the miss, $|\mu_{\Delta^+}/\mu_{\Delta^{++}}-\tfrac12|/\tfrac12$ in percent on a logarithmic axis, for the fixed O2 and O6 stencils and the per-state covariance-optimized estimators BLUE-O6, BLUE-O4 (the correlated linear fit in $B^2$) and BLUE-1 (the correlated constant fit); O4 and O8 lie with O6. The significance of the BLUE misses depends on whether the weights are reoptimized in the jackknife (Table~\ref{tab:decupletblue}).}
\label{fig:decuplet}
\end{figure*}

Table~\ref{tab:decuplet} and Fig.~\ref{fig:decuplet}(b) show the $\Delta^+$ result. The O2 estimate misses Eq.~(\ref{eq:deltaratio}) by 0.8--2.5\%, between 10 and 25 standard errors. O4, O6 and O8 reduce the violation by factors of 41--170 and satisfy the identity within $1.7\sigma$ at all six masses in both windows, with no cost in marginal error: the marginal errors of the $\Delta^+$ and $\Delta^{++}$ moments change by less than 1\% between O2 and O6. O8 remains consistent even though it uses the strongly mixed $8B$ point, because its weight on $y(8B)$ is $-1/2835$. A scan over 19 windows ($t_{\min}=8$--16, $t_{\max}=18$--22; listed in the reproducibility package) gives a similar picture: O2 is excluded in every window, by 3--25$\sigma$; O6 and O8 stay within $1.8\sigma$ at the five lighter masses and $2.6\sigma$ at the heaviest; O4, which leaves the $x^2$ term uncancelled, reaches $3.6\sigma$ in a few windows. The O6 residual is negative everywhere, but masses share configurations and adjacent windows share time slices, so this is a single tendency rather than many independent ones. The individual O6 moments shift with the window by up to $2.6\sigma$ (paired) at the lighter masses and $4.8\sigma$ at the heaviest, but the $\Delta^+$ and $\Delta^{++}$ shift together, so the ratio is far less window sensitive than the moments.

The $\Delta^0$ supplies an \emph{absolute} test: its weak-field moment is exactly zero, so any stencil output is pure higher-order leakage (Table~\ref{tab:delta0}). Expressed relative to the $\Delta^{++}$ moment on common configurations, O2 gives $-0.29$\% to $-0.52$\% (28--41$\sigma$) in the 10--18 window and $-0.46$\% to $-0.96$\% (15--37$\sigma$) in 12--20, O4 $-0.003$\% to $-0.010$\% (11--14$\sigma$) and $-0.008$\% to $-0.031$\% (12--15$\sigma$), and O6 at most $2\times10^{-6}$ in magnitude in the 10--18 window, within $1\sigma$ at the five lighter masses ($3.7\sigma$ at the heaviest) and $-1.8\times10^{-5}$ to $10^{-6}$ ($-3.6$ to $+0.4\sigma$) in the 12--20 window. At this precision, which reaches $10^{-6}$, small departures from the O6 family are resolved, consistent with the $\Delta^0$ O6 null test ($\chi^2_{\rm null}=14$--99 except at the heaviest mass in 10--18), but they are four orders of magnitude below the O2 leakage. O8, which gives weight to the strongly mixed $8B$ point, is displaced by 2.6--5.2$\sigma$ in the 10--18 window (0.7--3.5$\sigma$ in 12--20) at the same $10^{-6}$ level. The $\Delta^-$, which like the $\Delta^{++}$ has no nucleon partner, satisfies $\mu_{\Delta^-}/\mu_{\Delta^{++}}=-\tfrac12$ within $1.7\sigma$ for every stencil, including O2, at errors of $8\times10^{-5}$--$2\times10^{-3}$.

\subsection{Specification tests}

The tests distinguish the states. For the $\Delta^+$, O4 is rejected overwhelmingly ($\chi^2_{\rm null}=210$--320 in the production windows): the mixing term grows faster than $B^2$ in $y(B)$ between $4B$ and $8B$, which a model linear in $x$ cannot follow. O6 is not rejected in the 10--18 window ($\chi^2_{\rm null}=0.6$--2.6), is marginal in 12--20 (2.8--6.5), and fails in later windows, where the statistic grows above 100, as expected for a contribution that grows with Euclidean time. For the $\Delta^{++}$ both models are consistent with the data in the production windows ($\chi^2_{\rm null}\le4.9$ and 3.7), although the O6 statistic reaches 8.5 at the heaviest mass in the earliest scan window ($t=8$--18). Even where a test fails, the fixed stencils reproduce Eq.~(\ref{eq:deltaratio}): the tests diagnose the adequacy of a response family over all fields, while the stencil weights determine how much of a misspecification reaches the target.

\subsection{Covariance-optimized estimators}

For each state we form BLUE estimators, Eq.~(\ref{eq:blueweights}), from that state's noise-inclusive field-space covariance under three constraint sets: normalization only (BLUE-1, a correlated constant fit), the O4 constraints (BLUE-O4, a correlated linear fit in $B^2$), and the O6 constraints (BLUE-O6). Table~\ref{tab:decupletblue} compares them with fixed O6 on the same replicas. Fixed-weight errors apply the full-sample weights to every jackknife replica; reoptimized-weight errors re-derive the weights in each replica from the covariance of the remaining replicas. Both are delete-one errors; independent splits were done for the nucleon only (Sec.~\ref{sec:blue}).

\begin{table*}[t]
\caption{Covariance-optimized estimators on the $\Delta^+/\Delta^{++}$ control, window 10--18, with the temporal-covariance noise included throughout. $r-\tfrac12$ is in units of $10^{-4}$. Fixed-weight errors hold the full-sample weights fixed in every replica; reoptimized-weight errors re-derive the weights in each replica. Square brackets give the deviation in standard errors. $\sigma_\mathrm{BLUE}/\sigma_\mathrm{O6}$ is the ratio of the marginal errors of the $\Delta^+$ and $\Delta^{++}$ moments to those of the fixed O6 stencil. ``Joint'' rows use weights optimized for the ratio from the cross-state covariance; for them the last columns give the error of $r$ relative to fixed O6 (fixed and reoptimized weights). The per-state BLUE rows forgo the cross-state cancellation that identical weights inherit.}
\label{tab:decupletblue}
\centering
\setlength{\tabcolsep}{6pt}\begin{tabular}{cllllcc}
\toprule
& & & \multicolumn{2}{c}{Error on $r-\tfrac12$} & \multicolumn{2}{c}{$\sigma_\mathrm{BLUE}/\sigma_\mathrm{O6}$ ($\Delta^+$, $\Delta^{++}$)}\\
$m_\pi$ [MeV] & Estimator & $r-\tfrac12$ & Fixed $\bm w$ & Reoptimized & Fixed $\bm w$ & Reoptimized\\
\midrule
338 & O6 (fixed) & $-1.10$ & 0.94 [$-1.2$] & --- & --- & ---\\
    & BLUE-O6 & $-20.2$ & 30 [$-0.7$] & 25 [$-0.8$] & 0.98, 1.00 & 0.98, 1.01\\
    & BLUE-O4 & 431 & 46 [9.4] & 173 [2.5] & 0.96, 1.00 & 1.47, 1.01\\
    & BLUE-1 & 2384 & 115 [20.7] & 360 [6.6] & 0.87, 0.99 & 2.33, 1.03\\
    & joint O6 & 0.03 & 0.16 [0.2] & 0.16 [0.2] & \multicolumn{2}{c}{ratio: 0.17, 0.17}\\
    & joint O4 & 1.21 & 0.15 [8.1] & 0.20 [6.1] & \multicolumn{2}{c}{ratio: 0.16, 0.21}\\
\midrule
444 & O6 (fixed) & $-0.60$ & 0.54 [$-1.1$] & --- & --- & ---\\
    & BLUE-O6 & $-19.4$ & 16 [$-1.2$] & 14 [$-1.4$] & 0.99, 1.00 & 1.00, 1.01\\
    & BLUE-O4 & 327 & 30 [11.1] & 143 [2.3] & 0.97, 1.00 & 1.60, 1.01\\
    & BLUE-1 & 1657 & 62 [26.6] & 364 [4.6] & 0.92, 0.99 & 3.35, 1.07\\
    & joint O6 & 0.04 & 0.09 [0.4] & 0.09 [0.4] & \multicolumn{2}{c}{ratio: 0.16, 0.17}\\
    & joint O4 & 0.93 & 0.08 [11.6] & 0.11 [8.6] & \multicolumn{2}{c}{ratio: 0.15, 0.20}\\
\midrule
693 & O6 (fixed) & $-0.40$ & 0.26 [$-1.6$] & --- & --- & ---\\
    & BLUE-O6 & 2.5 & 3 [0.9] & 9 [0.3] & 1.00, 1.00 & 1.06, 1.03\\
    & BLUE-O4 & 148 & 20 [7.6] & 98 [1.5] & 0.98, 0.95 & 2.07, 1.00\\
    & BLUE-1 & 567 & 24 [23.5] & 265 [2.1] & 0.98, 0.94 & 4.70, 1.23\\
    & joint O6 & 0.07 & 0.04 [1.9] & 0.04 [1.8] & \multicolumn{2}{c}{ratio: 0.14, 0.15}\\
    & joint O4 & 0.45 & 0.03 [15.3] & 0.06 [7.7] & \multicolumn{2}{c}{ratio: 0.11, 0.23}\\
\bottomrule
\end{tabular}
\end{table*}

The ratio test compares two estimates whose fluctuations are almost perfectly correlated, so it is structurally favourable to any estimator that applies identical weights to both states, as the fixed stencils do; estimators optimized separately for each state forgo that cancellation. We therefore separate three comparisons.

\emph{Same constraints, misspecified family.} The O4 null test rejects the $\Delta^+$ ($\chi^2_{\rm null}=210$--320). Fixed O4 nevertheless satisfies the identity within $1.5\sigma$ at every mass, while BLUE-O4, the correlated linear fit in $B^2$ with the same constraints, violates it by 3--9\% in the 10--18 window and 10--17\% in 12--20 [Fig.~\ref{fig:decuplet}(c)], by 7.6--11$\sigma$ with fixed weights and 1.5--2.5$\sigma$ with reoptimized ones. For the $\Delta^+$ at the lightest mass, BLUE-O4 uses $\bm w=(18.9,-25.0,7.8,-0.68)$, with $L_4=-1.2\times10^3$ and $L_6=-1.5\times10^5$. That a rejected model gives a displaced answer is familiar \cite{DAgostini1994}; what the comparison adds is that the fixed stencil, which imposes the same model, is not displaced, because its leverage on the omitted modes is small. The correlated constant fit, BLUE-1, misses by 11--48\% in 10--18 and by 30--81\% in 12--20, but its model is rejected even more strongly, and its fixed counterpart, O2, also fails (by 0.8--2.5\%). With reoptimized weights, BLUE-1 and BLUE-O4 are also 1.5--4.7 times less precise than fixed O6 for the $\Delta^+$ moment in the 10--18 window (1.4--3.1 in 12--20), and their fixed-weight errors understate the reoptimized ones by factors of 3.1--11 (1.3--5.1 in 12--20). On the absolute control (Table~\ref{tab:delta0}) the same holds at O6 order, at a much smaller scale: in the 10--18 window the fixed O6 stencil gives zero within $1\sigma$ at the five lighter masses, while BLUE-O6 is displaced by 7.7--8.1$\sigma$ (5.9--10$\sigma$ in 12--20), by $2\times10^{-6}$--$2\times10^{-5}$ relative to $\mu_{\Delta^{++}}$, in a family that the $\Delta^0$ O6 null test rejects.

\emph{Accepted family, no misspecification.} For the $\Delta^{++}$, whose O4 test passes ($p=0.08$--0.57), BLUE-O4 differs from fixed O6 by at most 0.7\% and $1.5\sigma$.

\emph{Accepted family, per-state weights.} Within the accepted O6 family, optimizing each state separately gives no precision gain for either moment (marginal error ratios 0.98--1.06 in 10--18, 0.96--1.15 in 12--20), but it destroys the cross-state cancellation on which the identity's precision rests: in the 10--18 window the BLUE-O6 ratio error is 11--32 times that of fixed O6, and its deviation, $-20\times10^{-4}$ at the lighter masses, is within $1.4\sigma$ only because of that larger error. With the temporal covariance held fixed instead, the field-space covariance is an order of magnitude more ill-conditioned (for the $\Delta^+$, condition numbers $3\times10^5$--$4\times10^6$ against $4\times10^4$--$2\times10^5$), and BLUE-O6 then appears to reduce the $\Delta^+$ error by 28--41\% in the 10--18 window while violating the identity by 1.4--4.4\% there (up to 11\% in 12--20); that apparent gain was a property of the underestimated covariance. Optimizing the ratio itself, with weights chosen jointly from the cross-state covariance (``joint'' rows of Table~\ref{tab:decupletblue}), keeps the weights small and stencil-like, makes the ratio 4.3--6.8 times more precise than fixed O6 with reoptimized weights, and satisfies Eq.~(\ref{eq:deltaratio}) within $1.8\sigma$ with O6 constraints; with O4 constraints it violates it by 4.5--8.6$\sigma$, but by less than 0.06\%. Since the joint estimator is tuned to the statistic under test, this is not independent evidence. Judged on these controls, then, optimization within an accepted family neither helps nor measurably hurts; the large displacements arise where the family is misspecified, which the null-space test detects, or where per-state weights break a common-mode cancellation.

\emph{Regularization.} Shrinkage does not remove the effect. Shrinking each state's field-space covariance toward its diagonal \cite{LedoitWolf2004}, with intensities 0.1--0.9, reduces the BLUE-O4 leverage about fivefold but leaves its violation of the $\Delta^+$ identity at 9--13$\sigma$ at the lightest and heaviest masses; for BLUE-O6 it restores agreement with fixed O6. The data-driven (Ledoit--Wolf) intensity is only 0--0.04. That intensity is governed by a Frobenius-norm error, which the large eigenvalues dominate, so it does not mean that the smallest eigendirections, which BLUE exploits and which vary by a factor of 2--3 under resampling (Sec.~\ref{sec:blue_results}), are well determined. The synthetic study, however, shows BLUE-O4 displaced even with the true covariance: the displacement comes from the leverage of the optimum, not from noise in the estimated field-space covariance. On the $\Delta^0$, shrinkage worsens the precision of the optimized estimators by factors of 4--37 relative to fixed O6.

The control also shows when field-space cancellation can work. The mixing contamination enters $y(B)$ at a different order in $B$ than the target, so a stencil can remove it. For the nucleon, the same $N\to\Delta$ transition contributes to field-even energy shifts at $O(B^2)$. There it is a physical, paramagnetic part of the magnetic polarizability rather than a contamination; a field-space stencil cannot, and should not, remove it, and separating its contribution requires other information.

\section{Discussion}

The field-space viewpoint separates four questions often conflated in a single correlated fit: (i) the uncertainty of the estimated covariance; (ii) the statistical variance of a given estimator; (iii) its leverage on unconstrained response modes; and (iv) the size and physical interpretation of those modes. BLUE minimizes (ii) for a chosen constraint model and covariance. Additional configurations improve (i) but need not control (iii) or (iv), and (i) matters even for fixed stencils: the temporal-covariance noise is negligible for the moments but dominates the null contrasts.

\paragraph{Relation to practice.} In this language a correlated fit of a polynomial in $B^2$ (or, for charged states, in $|B|$) to all fields is a BLUE estimator with the corresponding constraints, and a fit with as many parameters as fields is a fixed stencil. Background-field calculations differ in which they use: single or weakest-field extractions, fits to subsets of fields, correlated polynomial fits in the field strength over several field quanta \cite{Chang2015LightNuclei}, correlated fits of a low-order polynomial with a higher-order term as a check \cite{Lee2005MagPol,Primer2014}, and fit families whose spread is taken as a systematic. Our results bear on the first two kinds only through the order of the stencil they imply; on the correlated-fit kind they say that the precision gain from optimization is small once the covariance noise is included, and that the result can be displaced in proportion to the leverage when the polynomial order is too low, while the null-space test, applied to the same data, flags that case. We have not re-analyzed published data, so we make no statement about specific published numbers.

\paragraph{Model uncertainty and robust design.} Bayesian model averaging over fit forms \cite{JayNeil2021} addresses the choice of truncation order by weighting models by their evidence; the null-space tests address the same choice by testing each family directly, and the two could be combined. The frontier of Sec.~\ref{sec:tau} is descriptive: a leverage is harmful only in proportion to the omitted coefficients it multiplies, since the displacement is $a_2L_4+a_3L_6+\cdots$. A prescriptive version would bound the omitted coefficients, for example by $\bm a^TC_a^{-1}\bm a\le1$, and minimize the worst-case mean-squared error $\bm w^T\Sigma\bm w+\bm w^T\Phi C_a\Phi^T\bm w$ (by the Cauchy--Schwarz inequality, $\bm w^T\Phi C_a\Phi^T\bm w$ is the largest squared bias $(\bm w^T\Phi\bm a)^2$ over that ellipsoid), where the columns of $\Phi$ are the omitted modes at the measured fields; this is minimax linear estimation \cite{Donoho1994} and bias-aware inference \cite{ArmstrongKolesar2018}. The diagnostics here supply its ingredients: the leverages, the resolved null contrasts that bound the omitted coefficients, and the noise-inclusive $\Sigma$. Pre-testing, as in Sec.~\ref{sec:synthetic}, is the simplest alternative, with the usual caveat that inference after model selection needs care.

\paragraph{Limitations.} All results come from one quenched ensemble with one field design and removed boundary-crossing links, at fields where $\ell_B$ is comparable to the source-to-boundary distance; the decuplet identities are exact only for valence-coupled fields with degenerate light quarks. The production windows are not free of excited states: two-state fits lower the magnitudes of the moments by 3--10\%, more than their statistical errors, and change the field-space covariance and with it the precision available to optimization (Sec.~\ref{sec:tempreg}). The upper ends of several ranges come from the heaviest mass, whose production fits are the most contaminated (Table~\ref{tab:windows}). The O6 specification test depends on the temporal analysis, and the two-state fits, whose gap is not stable in $t_{\min}$, do not settle which treatment is correct. For charged states the cancellation basis must reflect the physics of the response (Appendix~\ref{app:pion}). A field-even (polarizability) analysis of the same nucleon data requires spin-resolved energies, Eq.~(\ref{eq:spinavg}), and, for the proton, a stated gauge-origin convention; we leave it to separate work.

\section{Conclusions}

We have compared fixed cancellation stencils and covariance-optimized estimators of weak-field magnetic responses on archival nucleon and decuplet correlators, with contraction-level known-answer controls and a known-covariance synthetic study. The estimators are standard; the results concern their robustness.

First, the fixed $O(B^4)$--$O(B^8)$ stencils are nearly coincident and cost almost nothing in precision, because all fields share the same gauge configurations.

Second, weight-free null-space tests resolve departures from the O4 response family in most nucleon ensembles. Whether they resolve departures from the O6 family depends on the temporal analysis: with correlated temporal fits and the noise of the estimated temporal covariance included---which raises the error of the O6 contrast by a factor of 1.7--2.1 in a nested jackknife and 2.2--2.4 in full-pipeline bootstraps---they do not; with uncorrelated temporal fits they do at the heavier masses, also after the noise is included, but not at the lightest. Two-state temporal fits reject O4 for every ensemble and O6 for none, and they indicate that excited-state contamination accounts for most of the heaviest mass's outlying rejection in the production fits. The O6 verdict is therefore conditional on a temporal analysis whose systematics are of the same size as the effect tested.

Third, covariance optimization buys little precision: with the noise-inclusive covariance, 1--2\% for the nucleon moments at the five lighter masses and 7--8\% at the heaviest, none on independent halves of the data, and at most 8\% even if the estimated covariance were exact. These figures belong to the production single-exponential fits: with two-state fits the gain rises to 3--30\%, because the fitted excited-state term correlates the field responses, so the precision available to optimization is itself a property of the temporal analysis. Its weights carry three to four orders of magnitude more leverage on omitted response modes than the fixed stencils, and the optimized proton moments lie about one marginal standard error below fixed O4. Where the response family is misspecified, the optimized estimator is displaced in proportion to that leverage while the fixed stencil of the same order is not; a specification test applied first detects that case in our synthetic study, but has less power on the real data.

Fourth, the decuplet correlators supply targets known exactly from the lattice contractions. On $\mu_{\Delta^+}/\mu_{\Delta^{++}}=\tfrac12$, a higher-order response, plausibly field-induced mixing with the nucleon, displaces the lowest-order estimate by up to 25 standard errors. The fixed O4--O8 stencils remove it to within errors at the $10^{-4}$ level in the production windows, while the correlated linear fit in $B^2$, with the same constraints as O4, misses the identity by 3--17\% (the correlated constant fit by up to 81\%); the null-space tests reject the models of both fits. Within the accepted O6 family, per-state optimization neither gains precision nor measurably displaces the moments, but it breaks the cross-state cancellation on which the precision of the ratio test rests. On $\mu_{\Delta^0}=0$ the fixed O6 stencil reduces a 0.3--1\% leakage to the $10^{-6}$--$10^{-5}$ level.

With four fields, imposing $O(B^8)$ cancellation exhausts the available degrees of freedom, so additional field strengths, together with a worst-case design over a bounded set of omitted coefficients, are the natural route to combining higher-order control with statistical optimization.

\section*{Data and code availability}

The derived field-response vectors, the aligned jackknife samples (fixed and nested), the bootstrap and synthetic-study outputs, the field-space covariance matrices and full-precision optimized weights, and the scripts that generate every table and figure are archived at Zenodo, \href{https://doi.org/10.5281/zenodo.23869797}{doi:10.5281/zenodo.23869797}, and maintained at \url{https://github.com/scottmoe-oirg/constrained-field-space-paper1}. The underlying correlators are available from the authors upon reasonable request.

\begin{acknowledgments}
We thank Andrei Alexandru for writing the propagator and contraction code with which the correlators analyzed here were produced. This work was supported in part by the U.S. Department of Energy under Grant No.~DE-FG02-95ER40907. Most of the calculations were carried out at the DOE-sponsored National Energy Research Scientific Computing Center (NERSC). The authors used Claude (Anthropic) to assist with writing and debugging the analysis code, searching the literature, and drafting and revising the text; the authors directed the work, verified all output, and take full responsibility for the content.
\end{acknowledgments}

\appendix

\section{Archival reconstruction}
\label{app:archive}

\paragraph{Inventory.} The $\mathrm{poly11\mbox{-}small}$ proton archive contains 992 correlator files; the original analysis notebook reports 999, and seven proton configuration IDs beyond the common missing ID 675 are absent. The neutron archive contains 999 files. In the archived output the raw \texttt{MINUS} field tag corresponds to physical $+B$ and \texttt{PLUS} to physical $-B$; the $x$- and $y$-boundary flags of the production input are zero, and the boundary routine removes the links that would cross those boundaries before the Abelian phases are applied. Boundary-condition and action details not established for the other samples are not assumed to transfer between them.

\paragraph{Validation against the original analysis.} The original $\mathrm{poly0}$ proton notebook uses $t=8$--18. Reconstructing its 90-configuration double-ratio pipeline reproduces all 24 stored central field-fit exponents with maximum absolute discrepancy $1.3\times10^{-9}$. (The notebook fits the mean of the delete-one ratios; fitting the full-sample ratio instead changes the exponents by at most $2\times10^{-6}$.) For $\mathrm{poly11\mbox{-}small}$, the surviving proton and neutron files reproduce the archived field responses of the original analysis chain. All new field-space calculations use the surviving correlators rather than the saved notebook tables.

\paragraph{Normalization.} In the original nucleon notebooks, $m/(12\eta)$ is labeled as a natural-magneton or ``g-factor'' quantity and is subsequently multiplied by $0.938\,{\rm GeV}/M_N$; we find no intervening construction of the factor $2aM_N$ that would connect $m/(12\eta)=\mu/(ea)$ to that convention. We therefore use the explicitly derived Eq.~(\ref{eq:magresponse}). The difference is an overall rescaling of the moments.

\paragraph{Gauge origin.} In the production input each $y$-directed link carries the phase $\exp[i\,q\,\eta\,(x-x_0)]$, with $x_0=11$ for $\mathrm{poly11}$, at the source, and $x_0=0$ for $\mathrm{poly0}$. Moving the origin by $\delta=x_0-x_0'$ multiplies every $y$-link by the constant phase $e^{iq\eta\delta}$, which is the gauge transformation $\Omega_q(y)=e^{iq\eta\delta\,y}$. On a periodic $y$ direction $\Omega_q$ would have to satisfy $q\eta\delta L_y\in2\pi\mathbb Z$; with the boundary-crossing $y$-links removed there is no closing link, so $\Omega_q$ is an exact symmetry of the quark action for any $\delta$. Each propagator transforms as $S_q(x,y)\to\Omega_q(x)S_q(x,y)\Omega_q^\dagger(y)$, so a baryon correlator projected to zero spatial momentum acquires the factor $\exp[iQ\eta\delta\,(y_{\rm sink}-y_{\rm src})]$ inside the sink sum, where $Q$ is the baryon charge in units of the $d$-quark charge. For the neutron, $Q=0$ and the correlators are identical; for the proton the factor shifts the momentum of the projection. All 90 $\mathrm{poly0}$ configurations also occur in the 100-configuration sample with the $\mathrm{poly11}$ origin at the same quark masses (Sec.~\ref{sec:methods}), so the two origins can be compared configuration by configuration. The neutron magnetic and field-even responses agree to rounding. The paired proton O6 moments differ by only $-0.6$ and $-0.4$ jackknife standard errors at the lightest and heaviest masses, whereas the proton field-even energy shift is origin dependent (a $6.4\sigma$ paired difference at the heaviest mass), as expected for a charged state. Proton field-even quantities from this implementation therefore depend on the origin convention and are not used here.

\section{The decuplet identity at the level of the correlators}
\label{app:identity}

The decuplet two-point functions were computed with the routine \texttt{cfun\_decuplet} of the production code. With $\tilde S=C\gamma_3\,S\,C\gamma_3$ and a contraction kernel $F(S_1,S_2,\tilde S_3;\alpha,\beta)$ that is linear in its weights, the code evaluates
\begin{align}
C_{\Delta^{++}}&=F(S_u,S_u,\tilde S_u;\,2,4),\\
C_{\Delta^{+}}&=F(S_u,S_u,\tilde S_d;\,\tfrac23,\tfrac43)+F(S_d,S_u,\tilde S_u;\,\tfrac23,\tfrac43)\nonumber\\&\quad+F(S_u,S_d,\tilde S_u;\,\tfrac23,\tfrac43).
\end{align}
At zero field $S_u=S_d=S$ and the two are equal configuration by configuration. In the background field, $S_q=S+e_q\,\delta S+O(e_q^2)$. Let $D=\sum_{i=1}^3\partial_iF$ be the first-order variation of $F$ with all three slots varied, evaluated at $(S,S,\tilde S)$ with weights $(2,4)$. Then $\delta C_{\Delta^{++}}=e_u\,D$. In $C_{\Delta^+}$ each slot carries $e_d$ in one term and $e_u$ in the other two, so $\delta C_{\Delta^{+}}=\tfrac13(2e_u+e_d)\,D$. Hence
\begin{equation}
\frac{\delta C_{\Delta^{+}}(t)}{\delta C_{\Delta^{++}}(t)}=\frac{2e_u+e_d}{3e_u}=\frac12
\end{equation}
at every time slice, for every spin component and on every configuration. The $\Delta^0$ and $\Delta^-$ correlators are the analogous symmetric averages with two and three $d$ lines, so each slot carries $\tfrac13(e_u+2e_d)=0$ and $e_d=-\tfrac12e_u$ respectively, giving $\delta C_{\Delta^0}=0$ and $\delta C_{\Delta^-}=-\tfrac12\delta C_{\Delta^{++}}$ at first order. Because the zero-field spectral decompositions coincide, the coefficient of $t\,e^{-E_nt}$ in the first-order response, $-Z_n^2\,\delta E_n^{(1)}$, obeys the same relations for every non-degenerate level $n$.

Table~\ref{tab:identitycheck} tests this on the data through the field-odd parts $O_X(t)=C_X(+B)-C_X(-B)$.

\begin{table}[h]
\caption{Deviation $|O_{\Delta^+}/O_{\Delta^{++}}-\tfrac12|$: median over the 892 configurations and both spin components at each mass, then median over the six masses. The deviation scales as $f^2$ and vanishes as $B\to0$; its coefficient grows with Euclidean time.}
\label{tab:identitycheck}
\centering
\begin{tabular}{rcccc}
\toprule
& $f=1$ & $f=2$ & $f=4$ & $f=8$\\
\midrule
$t=5$  & $7.5\times10^{-4}$ & $3.0\times10^{-3}$ & $1.2\times10^{-2}$ & $4.4\times10^{-2}$\\
$t=10$ & $4.7\times10^{-3}$ & $1.9\times10^{-2}$ & $7.6\times10^{-2}$ & 0.31\\
$t=14$ & $9.9\times10^{-3}$ & $4.0\times10^{-2}$ & 0.17 & 0.73\\
$t=18$ & $1.3\times10^{-2}$ & $5.4\times10^{-2}$ & 0.23 & 1.03\\
\bottomrule
\end{tabular}
\end{table}

\section{Charged-particle response and a nonanalytic stencil}
\label{app:pion}

A charged spin-zero state illustrates why the cancellation basis must follow the physics. Its field-even response may contain
\begin{equation}
\Delta E(B)=a|B|+bB^2+c|B|^3+dB^4+\cdots,
\label{eq:landau}
\end{equation}
so that $\Delta E(B)/B^2=a/|B|+b+c|B|+dB^2+\cdots$. Moreover, expanding the charged-particle Landau energy can itself generate analytic $B^2$ contributions. Consequently an analytic even-power cancellation hierarchy is only a diagnostic of the assumed basis here, not a determination of a physical polarizability. Landau-level projection \cite{TiburziVayl2013,Bignell2020Pion} and four-point-function methods, which avoid the background-field expansion altogether~\cite{Lee2023PionElec4pt,Lee2023PionMag4pt,LeeNeutral:2024,Wang:2024}, address this directly.

For a charged state, evenness under field reversal does not guarantee analyticity in $B^2$. As an illustration, preserving a constant while cancelling $1/|B|$, $|B|$, and $B^2$ at $B,2B,4B,8B$ yields
\begin{equation}
\bm w=\left(-\frac83,\frac{22}{3},-\frac{13}{3},\frac23\right).
\end{equation}
This is a Laurent-series stencil in $|B|$. Its weights are large and alternating, but because they still sum to one, a fluctuation common to all fields passes through unamplified, so its cost depends on the cross-field correlation of the channel, and an exact physics-informed nulling prescription can be statistically fragile when that correlation is weak. Whether a $1/|B|$ term is needed at all is itself a physical question: the magnetic length, $\ell_B=(|e|B)^{-1/2}\simeq 15.6a$ at the weakest field and $5.5a$ at the strongest, is comparable to the 24-site box with its boundary-crossing links removed, so Landau levels need not have formed. Predicting or subtracting a known Landau contribution before fitting may be preferable to forcing its cancellation when the cross-field correlation is weak. More generally, \emph{the stencil must reflect the physics of the field response}, and field placement is itself part of estimator design.

\bibliography{field_space_paper1_refs}
\end{document}